\documentclass[a4paper,11pt]{article}
\usepackage{t1enc}
\usepackage{amsmath,bm}
\usepackage{amssymb}
\usepackage{enumerate}
\usepackage{fullpage}
\usepackage{verbatim}
\usepackage{graphicx}
\usepackage{psfrag}

\usepackage{hyperref}
\usepackage{authblk}

\newcommand{\begy}{\bar{1}}

\newcommand{\On}[1]{\mathrm{O}(#1)}
\newcommand{\hth}{\hat{\theta}}
\newcommand{\halpha}{\hat{\alpha}}
\newcommand{\hbeta}{\hat{\beta}}
\newcommand{\hgamma}{\hat{\gamma}}

\newcommand{\kv}{\underline{k}_{12}}
\newcommand{\kvtt}{\underline{k}^{t_1t_2}_{12}}
\newcommand{\kvt}[1]{\underline{k}^{t_{#1}}_{12}}

\newcommand{\Am}{\mathbf{A}}
\newcommand{\alpham}{\bm{\alpha}}
\newcommand{\halpham}{\hat{\bm{\alpha}}}
\newcommand{\Xm}{\mathbf{X}}
\newcommand{\Ym}{\mathbf{Y}}

\newcommand{\Fvec}{\underline{\mathcal{F}}}
\newcommand{\Bv}{\underline{B}}
\newcommand{\Cv}{\underline{C}}

\newcommand{\betav}{\underline{\beta}}
\newcommand{\gammav}{\underline{\gamma}}
\newcommand{\hbetav}{\hat{\underline{\beta}}}
\newcommand{\hgammav}{\hat{\underline{\gamma}}}

\newcommand{\ui}[2]{u_{#2}^{({#1})}}
\newcommand{\vi}[2]{v_{#2}^{({#1})}}
\newcommand{\wi}[2]{v_{#2}^{({#1})}}
\newcommand{\as}{\alpha}
\newcommand{\ad}{\dot{\alpha}}

\newcommand{\tr}[1]{\mathrm{tr}_{#1}}

\newcommand{\ket}[1]{|#1\rangle}
\newcommand{\bra}[1]{\langle #1|}

\title{Nonstandard Bethe Ansatz equations for open O(N) spin chains}
\author[1,2]{\small Tam\'as Gombor}
\affil[1]{\footnotesize MTA Lend\"ulet Holographic QFT Group, Wigner Research Centre for Physics,\protect\\Konkoly-Thege Mikl\'os u. 29-33, 1121 Budapest , Hungary}
\affil[2]{\footnotesize Institute for Theoretical Physics, Roland E\"otv\"os University,\protect\\1117 Budapest, P\'azm\'any P\'eter s\'et\'any 1/A, Hungary}
\date{}

\begin{document}
\maketitle

\abstract{
  The double row transfer matrix of the open $\mathrm{O}(N)$ spin chain is diagonalized and the Bethe Ansatz equations are also derived by the algebraic Bethe Ansatz method including the so far missing case when the residual symmetry is $\mathrm{O}(2M+1)\times\mathrm{O}(2N-2M-1)$.
  In this case the boundary breaks the ''rank'' of the $\mathrm{O}(2N)$ symmetry leading to nonstandard Bethe Ansatz equations in which the number of Bethe roots is less than as it was in the periodic case.
  Therefore these cases are similar to soliton-nonpreserving reflections.
}


\section{Introduction}

A number of methods have been developed in the past for the calculation of the spectrum of quantum-integrable systems including coordinate, algebraic and analytic Bethe-Ansatz.
For the algebraic and analytic Bethe-Ansatz, the \emph{quantum R-matrix}, which satisfies the \emph{Yang-Baxter equation}, is indispensable.
In this paper we will deal with quantum spin chains with a simple symmetry group ($G$) which have rational R-matrix.
It is known that most of these types of R-matrices are related to some 1+1 dimensional relativistic quantum field theories which are integrable and the two particle scattering matrices are proportional to their R-matrices.

For the integrability of open spin-chains, the existence of the R-matrix is not sufficient, we also need a \emph{K-matrix}.
The K-matrix must satisfy the \emph{boundary Yang-Baxter equation} or otherwise the $KRKR = RKRK$ equation.
For rational R-matrices, the unit matrix can be a K-matrix and in this case the symmetry of the system is not broken by the boundary.
For other solutions of the boundary Yang-Baxter equation, the K-matrix breaks the bulk symmetry group $G$.
These K-matrices are classified in \cite{Arnaudon:2004sd,Arnaudon:2003gj,Doikou:2000yw,Moriconi:2001xz,MacKay:2001bh}.
For these K-matrices, we can find 1+1 integrable QFTs that have boundary scattering matrices which are proportional to the above K-matrices.
If the residual symmetry group is $H$, it has previously been found that the $G/H$ quotients are symmetric spaces for all possible K-matrices.

The possible integrable boundaries can be divided into two groups.
The first group contains those for which $\mathrm{rank}(\mathfrak{h}) = \mathrm{rank}(\mathfrak{g})$, where $\mathfrak{h}$ and $\mathfrak{g}$ are the Lie algebras of the $H$ and $G$ groups.
In these cases the eigenvalue of the transfer matrix and the Bethe-Ansatz equations are known and they have the same structure as in the periodic case (they have the same type of Bethe roots).

For the second case we have $\mathrm{rank}(\mathfrak{h}) < \mathrm{rank}(\mathfrak{g})$.
One such case is the so-called \emph{soliton-nonpreserving reflection} \cite{Doikou:2000yw}.
For such systems, transfer matrix eigenvalues and Bethe-Ansatz equations were determined by the analytic Bethe-Ansatz \cite{Arnaudon:2005rq,Arnaudon:2004sd}.
Recently, this model was solved by algebraic Bethe Ansatz method. \cite{Gerrard:2017igy}.
These open spin-chains can be related to some SU(N) principal chiral field theories \cite{Short:2002yr,MacKay:2001bh}.
In addition, there is a non-investigated case when $G=\On{2N}$ and $H=\On{2M+1}\times\On{2N-2M-1}$.

The purpose of the article is to develop an algebraic Bethe-Ansatz method by which the Bethe Ansatz equations of this so far missing case can be determined.
For the periodic $\On{2N}$ spin chains there are two nested algebraic Bethe-Ansatz methods which can be used for the diagonalization of the transfer matrices.
The first one was developed in \cite{deVega:1986xj}.
The basis of this method is that the problem is returned to diagonalization of an $\mathrm{SU}(N)$ symmetric transfer matrix in the first step of the nesting.
It is possible because the R-matrix of the $\On{2N}$ model has a six-vertex block-form.
This method was generalized for open spin chains in \cite{Gombor:2015kdu}. 
Unfortunately, that procedure cannot be used for the model with $\On{2M+1}\times\On{2N-2M-1}$ symmetric boundary because its K-matrix is not block diagonal.
 
The second method was developed in \cite{Martins:1997wb}.
In this method we have to diagonalize a $\On{2N-2k}$ transfer matrix at the $k$th step of the nesting.
In this paper we develop the generalization of this method for open spin chains.
We will see that the method can be used for the diagonalization of the transfer matrix with $\On{2M+1}\times\On{2N-2M-1}$ symmetric K-matrix as well.

The paper is structured as follows:
The next section summarizes the possible integrable reflections.
We will find that all $H$ are maximal subgroups of $G$.
We briefly describe the classification of maximal subgroups by regular and special subgroups.
We will see that the regular maximal subgroups correspond to symmetry rank preserving reflexions.

In the third section the algebraic Bethe-Ansatz is described for open $\On{2N}$ spin chains, which is the generalization of the periodic case \cite{Martins:1997wb}.
Using this we construct Bethe-Ansatz equations for all $\On{2N}$ K-matrices including the $H=\On{2M+1}\times\On{2N-2M-1}$ case which was not studied previously.
This Bethe-Ansatz method can be applied to the $\On{2N+1}$ type spin chains and the results are summarized in the first appendix.

In the final section we compare the non-soliton-preserving reflections of the $\mathrm{SU}(4)$ spin chains and the symmetry rank breaker $H=\On{5},\On{3}\times\On{3}$ reflections of the $\On{6}$ spin chain.
We will see that the $\On{6}$ reflections can be obtained by fusion in the $\mathrm{SU}(4)$ model.

\section{Connection between maximal subalgebras and reflection matrices}

In this section we overview the known solutions of the \emph{boundary Yang-Baxter equation (BYBE)} for spin chains which have a symmetry group with simple Lie-algebra $\mathfrak{g}$.
We ignore the exceptional ones.
We will use the following conventions for the fundamental representation of the $R-matrices$:
\begin{itemize}
  \item $\mathfrak{g}=\mathfrak{su}(n)$ case:
	\[
	  R_{ab}^{cd}(u) = \delta_a^c \delta_b^d - \frac{2}{u} \delta_a^d \delta_b^c 
	\]

  \item $\mathfrak{g}=\mathfrak{so}(n)$ case:
	\[
	  R_{ab}^{cd}(u) = \delta_a^c \delta_b^d - \frac{2}{u} \delta_a^d \delta_b^c - \frac{2}{n-2-u} \delta_{ab} \delta^{cd} 
	\]
  
  \item $\mathfrak{g}=\mathfrak{sp}(n)$ case ($n$ is even):
	\[
	  R_{ab}^{cd}(u) = \delta_a^c \delta_b^d - \frac{2}{u} \delta_a^d \delta_b^c - \frac{2}{n+2-u} U_{ab} U^{cd} 
	\]
\end{itemize}
where
\[
		  U = \begin{pmatrix}
			0 & \mathbb{I}_{n/2} \\
			-\mathbb{I}_{n/2} & 0
		  \end{pmatrix}.
\]
The reflection matrices have residual symmetry with Lie-algebra $\mathfrak{h}$ which is a subalgebra of $\mathfrak{g}$.

There are four kinds of $\mathfrak{g}$:
\begin{enumerate}
  \item $\mathfrak{g}=\mathfrak{su}(n)$
	In this case there are three kind of known solution.
	\begin{enumerate}
	  \item $\boxed{\mathfrak{h} = \mathfrak{su}(k) \oplus \mathfrak{su}(n-k) \oplus \mathfrak{u}(1)}$ 
		where $1<k\le n/2$ 
		and $\boxed{\mathfrak{h} = \mathfrak{su}(n-1) \oplus \mathfrak{u}(1)}$

		These reflection matrices can be diagonalized with $\exp(\mathfrak{g})$ transformation.
		In that form the matrix contains one free parameter, $c$.

		\[
			K(u) = \begin{pmatrix}
				(c+u)\mathbb{I}_k & 0 \\
				0 & (c-u)\mathbb{I}_{n-k} 
			\end{pmatrix}
		\]

		For this reflection $\mathrm{rank}(\mathfrak{g})=\mathrm{rank}(\mathfrak{h})$.
		\label{An_a}
	  
	  \item $\boxed{\mathfrak{h} = \mathfrak{so}(n)}$

		This case belongs to a representation changing reflection where a particle goes to its anti-particle.
		If we do not want to work with direct sum representation we have to use this reflection in the real reps of $\mathfrak{su}(n)$.
		This reflection has not got any free parameter.
		
		For this reflection $\mathrm{rank}(\mathfrak{g})>\mathrm{rank}(\mathfrak{h})$.
		\label{An_b}

	  \item $\boxed{\mathfrak{h} = \mathfrak{sp}(n)}$ if $n$ is even
		
		This case also belongs to a representation changing reflection.
		In this case if we choose a basis in the particle's and the anti-particle's space the reflection matrix can be written in the following form:
		\[
		  K(u) = \begin{pmatrix}
			0 & \mathbb{I}_{n/2} \\
			-\mathbb{I}_{n/2} & 0
		  \end{pmatrix}.
		\]
		
		For this reflection $\mathrm{rank}(\mathfrak{g})>\mathrm{rank}(\mathfrak{h})$.
		\label{An_c}
	\end{enumerate}

  \item $\mathfrak{g}=\mathfrak{so}(2n+1)$ 
	\begin{enumerate}
	  \item $\boxed{\mathfrak{h} = \mathfrak{so}(2k) \oplus \mathfrak{so}(2n+1-2k)}$
		where $1 < k< n-1$ 
		and $\boxed{\mathfrak{h} = \mathfrak{so}(2n)}$

		After we diagonalized this reflection matrix there is no free parameter.
		\[
			K(u) = \begin{pmatrix}
				(n+1/2-2k+u)\mathbb{I}_{2k} & 0 \\
				0 & (n+1/2-2k-u)\mathbb{I}_{2n+1-2k} 
			\end{pmatrix}
		\]
		
		For this reflection $\mathrm{rank}(\mathfrak{g})=\mathrm{rank}(\mathfrak{h})$.
		\label{Bn_a}

	  \item $\boxed{\mathfrak{h} = \mathfrak{so}(2n-1) \oplus \mathfrak{u}(1)}$
		
		This case contains one free parameter.
		\[
			K(u) = \begin{pmatrix}
				(n-3/2)^2-c^2-u^2 & 2icu & 0 \\
				-2icu & (n-3/2)^2-c^2-u^2 & 0 \\ 
				0 & 0 & ((n-3/2-u)^2-c^2)\mathbb{I}_{2n-1}
			\end{pmatrix}
		\]
		
		For this reflection $\mathrm{rank}(\mathfrak{g})=\mathrm{rank}(\mathfrak{h})$.
		\label{Bn_b}

	\end{enumerate}

  \item $\mathfrak{g}=\mathfrak{sp}(2n)$
	\begin{enumerate}
	  \item $\boxed{\mathfrak{h} = \mathfrak{sp}(2k) \oplus \mathfrak{sp}(2n-2k)}$
		where $1< k \le n/2$

		In this case there is no free parameter.
		\[
			K(u) = \begin{pmatrix}
				(n-2k+u)\mathbb{I}_{k} & 0 & 0 & 0 \\
				0 & (n-2k-u)\mathbb{I}_{n-k} & 0 & 0 \\ 
				0 & 0 & (n-2k+u)\mathbb{I}_{k} & 0 \\
				0 & 0 & 0 & (n-2k-u)\mathbb{I}_{n-k} 
			\end{pmatrix}
		\]
		
		For this reflection $\mathrm{rank}(\mathfrak{g})=\mathrm{rank}(\mathfrak{h})$.
		\label{Cn_a}

	  \item $\boxed{\mathfrak{h} = \mathfrak{su}(n) \oplus \mathfrak{u}(1)}$
		
		There is also  one free parameter.
		\[
			K(u) = \begin{pmatrix}
				c\mathbb{I}_{n} & iu\mathbb{I}_{n} \\
				-iu\mathbb{I}_{n} & c\mathbb{I}_{n}  
			\end{pmatrix}
		\]

		For this reflection $\mathrm{rank}(\mathfrak{g})=\mathrm{rank}(\mathfrak{h})$.
		\label{Cn_b}

	\end{enumerate}

  \item $\mathfrak{g}=\mathfrak{so}(2n)$
	where $n>2$ (We ignore the $\mathfrak{so}(4)$ case because it is not a simple Lie-algebra)
	\begin{enumerate}
	  \item $\boxed{\mathfrak{h} = \mathfrak{so}(2k) \oplus \mathfrak{so}(2n-2k)}$
		where $1< k \le n/2$

		This case is similar to case \ref{Bn_a}.  
		\[
			K(u) = \begin{pmatrix}
				(n-2k+u)\mathbb{I}_{2k} & 0 \\
				0 & (n-2k-u)\mathbb{I}_{2n-2k} 
			\end{pmatrix}
		\]
		\label{Dn_a}
		
	  \item $\boxed{\mathfrak{h} = \mathfrak{so}(2n-2) \oplus \mathfrak{u}(1)}$

		This case is similar to \ref{Bn_b}. 
		\[
			K(u) = \begin{pmatrix}
				(n-2)^2-c^2-u^2 & 2icu & 0 \\
				-2icu & (n-2)^2-c^2-u^2 & 0 \\ 
				0 & 0 & ((n-2-u)^2-c^2)\mathbb{I}_{2n-2}
			\end{pmatrix}
		\]
		\label{Dn_b}
		
	  \item $\boxed{\mathfrak{h} = \mathfrak{su}(n) \oplus \mathfrak{u}(1)}$
		
		It has one free parameter.
		\[
			K(u) = \begin{pmatrix}
				c\mathbb{I}_{n} & iu\mathbb{I}_{n} \\
				-iu\mathbb{I}_{n} & c\mathbb{I}_{n}  
			\end{pmatrix}
		\]

		For this reflection  $\mathrm{rank}(\mathfrak{g})=\mathrm{rank}(\mathfrak{h})$.
		\label{Dn_c}

	  \item $\boxed{\mathfrak{h} = \mathfrak{so}(2k+1) \oplus \mathfrak{so}(2n-2k-1)}$
		where $1\le k <n/2$
		and $\boxed{\mathfrak{h} = \mathfrak{so}(2n-1)}$

		\[
			K(u) = \begin{pmatrix}
				(n-2k-1+u)\mathbb{I}_{2k+1} & 0 \\
				0 & (n-2k-1-u)\mathbb{I}_{2n-2k-1} 
			\end{pmatrix}
		\]
		This case is similar as \ref{Bn_a} and \ref{Dn_a} but there is a big difference because the $\mathrm{rank}(\mathfrak{h})=\mathrm{rank}(\mathfrak{g})-1$ so the K-matrix breaks the rank of the symmetry algebra.
		\label{Dn_d}

	\end{enumerate}

\end{enumerate}

We can make some interesting comments. 
The $\mathfrak{h}$s are \emph{maximal} subalgebras of $\mathfrak{g}$s for all solution.
We can first observe that the number of free parameters in the reflections are equal to the number of $\mathfrak{u}(1)$s in $\mathfrak{h}$.
In addition, one can divide the reflections into two groups:\begin{itemize}
	\item \ref{An_a}, \ref{Bn_a}, \ref{Bn_b}, \ref{Cn_a}, \ref{Cn_b}, \ref{Dn_a}, \ref{Dn_b}, \ref{Dn_c}

	In these cases the symmetry of the boundary has the same rank as the bulk.
	These are the \emph{regular reflections} and we will denote them by \emph{R-reflections}.

	\item \ref{An_b}, \ref{An_c}, \ref{Dn_d}

	In these cases the symmetry of the boundary has lower rank than the bulk.
	These are the \emph{non-regular reflections} or \emph{special reflections} and we will denote them by \emph{S-reflections}.
\end{itemize}

The Bethe-Ansatz equations for the R-reflections are well known \cite{Arnaudon:2003gj,2005PhLA}.
In these cases the calculations and results are very similar to the periodic case.
However the known BA equations for S-reflections are very different \cite{Arnaudon:2004sd,Gerrard:2017igy}.
There is one S-reflection whose BA equations have not been calculated yet.
This is the case \ref{Dn_d}.
In the next section we will derive the BA equations for that case using algebraic Bethe-Ansatz method.

Let us go through the classification of the maximal subalgebras of the simple Lie-algebras.

We will use two class of maximal subalgebras:
\begin{itemize}
	\item a \emph{regular maximal subalgebra} is that whom Cartan-subalgebra is the same as that of the original algebra. We will denote them by \emph{R-subalgebras}.
	\item All the other maximal subalgebras are called \emph{special} and we will denote them by \emph{S-subalgebras}.
\end{itemize}

The R-subalgebras can simply be obtained by using the Dynkin diagrams but it is not easy to find S-subalgebras.

\begin{figure}
	\caption{The extended Dynkin-diagrams.}
	\centering
	\psfrag{t}{$\theta$}
	\psfrag{An}{\Large $\tilde{A}_n$}
	\psfrag{Bn}{\Large $\tilde{B}_n$}
	\psfrag{Cn}{\Large $\tilde{C}_n$}
	\psfrag{Dn}{\Large $\tilde{D}_n$}
	\includegraphics[width=0.7\textwidth]{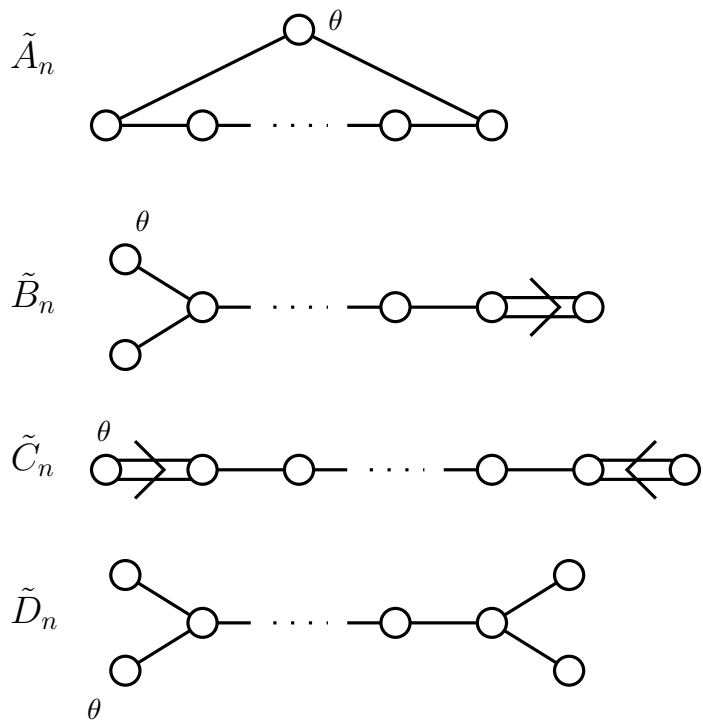}
	\label{fig:extended}
\end{figure}

There are two kinds of R-subalgebras: semi-simple and not semi-simple.
First, let's look at the semi-simple ones.
To find them the trick is to draw the extended Dynkin diagrams (Figure \ref{fig:extended}) which is obtained by adding the most negative root $\theta$ to the simple roots.
If we erase one root (except for the most negative one) from this diagram we get semi-simple R-subalgebra or the original simple algebra. For example if we remove the kth root from the $\tilde{C}_n$ diagram then we get a $C_k\oplus C_{n-k}$ semi-simple algebra or if any root is removed from the $\tilde{A}_n$ diagram then the original $A_n$ algebra is returned.

Now let us see how we can get non semi-simple subalgebras.
It has been mentioned earlier that there are roots in the extended diagram that by removing any of them the original algebra can be recovered.
If we erase the same root from the non-extended Dynkin-diagram we get a semi-simple algebra plus a $\mathfrak{u}(1)$ factor.
For example if we remove the kth root from the $A_{n-1}$ diagram then the remaining diagram will correspond to $A_{k-1}$ plus  $A_{n-k-1}$.
That is, the $\mathfrak{su}(n)$ algebra has $\mathfrak{su}(k) \oplus \mathfrak{su}(n-k) \oplus \mathfrak{u}(1)$ subalgebra.

Based on these, we can determine all the regular subalgebras.
From the definition it follows that the R-subalgebras have the same rank as the original algebra.

It is then easy to check the statement at the beginning of the section for regular reflections and subalgebras i.e. all the R-algebras have a R-reflections.
For example, if we take the $\mathfrak{g}=\mathfrak{sp}(2n)$ case and delete an internal point on the $\tilde{C}_n$ diagram we get a $C_k$ plus $C_{n-k}$ diagram which belongs to the reflection \ref{Cn_a}.
If we erase the last root we get the original $C_n$.
So if we remove this node from the original diagram we get a non semi-simple maximal subalgebra $\mathfrak{sp}(2n-2) \oplus \mathfrak{u}(1)$ which belongs to the reflection $\ref{Cn_b}$. 
So we can see the one-to-one correspondence in this $\mathfrak{g}=\mathfrak{sp}(2n)$ case.
The same can be checked in the other cases.

Let us continue by examining S-reflections.
It is easy to check that $\mathfrak{so}(n) \subset \mathfrak{su}(n)$, $\mathfrak{sp}(n) \subset \mathfrak{su}(n)$ and $\mathfrak{so}(2k+1) \oplus \mathfrak{so}(2n-2k-1) \subset \mathfrak{so}(2n)$ subalgebras are maximals but not regulars so they are S-subalgebras.
So we can see that all S-reflections have S-subalgebra symmetry.

\section{Algebraic Bethe-Ansatz for the $\On{2N}$ model}

In this section we generalize the method of \cite{Martins:1997wb} for open spin chains.
Because of the complexity of the commutation relations of the elements of the transfer matrix we do not derive precisely the cancellation of unwanted terms.
To check the correctness of the results we can compare the results of the subsections \ref{subsec:ABA}, \ref{subsec:UN}, \ref{subsec:u1} with the previous results \cite{Gombor:2015kdu,Arnaudon:2003gj}.
We then can assume that this method gives the correct result for the \ref{Dn_d} case as well.

\subsection{Conventions}

We will use the following convention for the R-matrix:
\[
  R_{12}(u) = I_{12} -\frac{2}{u}P_{12} -\frac{2}{\hat{u}}K_{12} = I_{12} +d(u)P_{12} +e_N(u)K_{12}, 
\]
where $\hat{u}:=2N-2-u$ and $1$ and $2$ denote two copies of a $2N$ dimensional vector space.
We want to diagonalize the double row transfer matrix:
\[
  D(\theta) = \tr{0} \left[  K_0^L(\hth) M_0(\theta) \right],
\]
where $M_0$ is the double row monodromy matrix
\[
  M_0(\theta) = T_0(\theta) K_0^R(\theta) \hat{T}_0(\theta),
\]
where $T_0$ and $\hat{T}_0$ are two one row monodromy matrices
\begin{align*}
  T_0(\theta) &= R_{01}(\theta) \cdots R_{0L}(\theta), &
  \hat{T}_0(\theta) &= R_{L0}(\theta) \cdots R_{10}(\theta).
\end{align*}
From earlier results we know the Bethe Ansatz equations of the $\On{2N}$ model with regular boundary conditions \cite{Gombor:2015kdu}.
The result can be easily illustrated.

First look at the $D_N$ Dynkin-diagram.
After that we mark the simple root ($M$ or $1$ or $+$) which does not belong to the symmetry algebra of the boundary.
We can assign massless particles to the nodes of the Dynkin-diagram.
The type $M$ particle will get a nontrivial boundary reflection factor because the simple root of these particles do not commute with the reflection matrix.
\begin{center}
\psfrag{0}{$0$}
\psfrag{1}{$1$}
\psfrag{2}{$2$}
\psfrag{L}{$M$}
\psfrag{Nm2}{$N-2$}
\psfrag{p}{$+$}
\psfrag{m}{$-$}
\psfrag{4a}{\large 4a}
\psfrag{4b}{\large 4b}
\psfrag{4c}{\large 4c}
\includegraphics[width=0.6\textwidth]{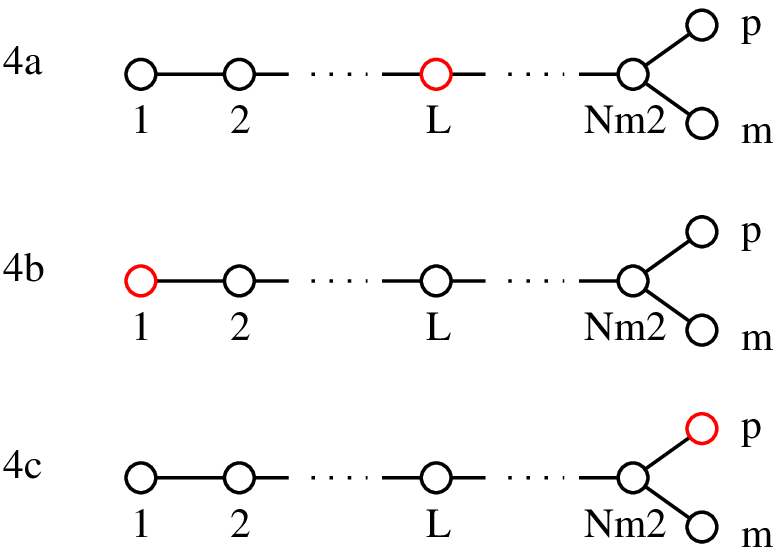}
\end{center}

If we know the scattering and reflection phases of these magnons we also know the BAEs. Only the connected nodes have non-trivial scattering factors:
\begin{align*}
  s_{ij}(u) &= \frac{u+\alpha_i\cdot\alpha_j}{u-\alpha_i\cdot\alpha_j}.
 \end{align*}
 where the $\alpha_i$s are the usual simple roots of the $D_N$ algebra.
\[
	\alpha_i \cdot \alpha_j = 
	\begin{pmatrix}
		 2 & -1 &    &        &    &    & \\
		-1 &  2 & -1 &        &    &    & \\
		   & -1 &  2 &        &    &    & \\
		   &    &    & \ddots &    &    & \\
		   &    &    &        &  2 & -1 & \\
		   &    &    &        & -1 &  2 & -1 &    &    \\
		   &    &    &        &    & -1 &  2 & -1 & -1 \\
		   &    &    &        &    &    & -1 &  2 &    \\
		   &    &    &        &    &    & -1 &    &  2 \\
	\end{pmatrix}
\]

The non-trivial reflection factor is the following:
\begin{align*}
  r_M(u) &= \frac{N-M+u}{N-M-u}, & r_1(u) &= \frac{N-1-c+u}{N-1-c-u}, & r_+(u) &= \frac{N-1-c+u}{N-1-c-u}.
\end{align*}

In the next subsection we will present a different procedure which can be used for the calculation of this result.
This calculation is the generalization of the one in \cite{Martins:1997wb} to open spin chains.

\subsection{The nesting for $R$-reflections} \label{subsec:nesting}
In this subsection, we describe the nested Algebraic Bethe Ansatz for open spin chains.
In the zero step the diagonalization of the transfer matrix of the SO($2N$) model is returned to the diagonalization of SO($2N-2$) transfer matrix.
In the first step the case SO($2N-2$) is returned to an SO($2N-4$) case, and so on.
In the last step we have to solve an SO($4$) model.
At this step the SU(2)$\times$SU(2) base is used because the R-matrix is factored at this base.
The K-matrices, which are examined in this subsection, are also factorized.
Non-factorizing K-matrices are investigated in the following subsection.

At the nested Bethe Ansatz of the periodic case we saw there were two types of ''unwanted'' terms: the traditional (in which the rapidities are interchanged) or the so-called ''easy unwanted'' terms which include not only a creation operator but also a annihilation operator of the pseudo-vacuum \cite{Martins:1997wb}.

These ''easy unwanted'' terms can be omitted in one-magnon states because they annihilate the pseudo-vacuum.
However at the multi-particle states these can not be omitted.
For the disappearance of this ''easy unwanted'' terms it was necessary to define n-particle operators with a complicated recursion formula.
We have seen at the periodical case that the formula of the n-particle state can be derived by an alternative method.
At this method we required that the eigenvectors must be symmetrical in rapidities.
We will use this second method to derive the formula for n-particle eigenstates.

\subsubsection{The zeroth step}
At the begin of every step of nesting we have to choose a new basis: $\{\ket{1},\ket{2},\ket{3},\dots,\ket{2N}\}\longrightarrow\{\ket{1}_c,\ket{3},\dots,\ket{2N},\ket{\bar{1}}_c\}$ where
\begin{align*}
	\frac{1}{\sqrt{2}}\left(\ket{1}+i\ket{2}\right) \longrightarrow \ket{1}_c,\\
	\frac{1}{\sqrt{2}}\left(\ket{1}-i\ket{2}\right) \longrightarrow \ket{\bar{1}}_c.
\end{align*}
So the $2N$ dimensional spaces are decomposed to the next direct sum: $\mathbb{C}^{2N}=\mathbb{C}\oplus\mathbb{C}^{2N-2}\oplus\mathbb{C}$.
The R-matrix in this new basis can be written in the following way:
\begin{multline*}
  R_{12}(u) =\\ \left(
  \begin{array}{ccc|ccc|ccc}
	a(u) & 0 & 0 & 0 & 0 & 0 & 0 & 0 & 0\\
	0 & I_2 & 0 & d(u)\kvt2 & 0 & 0 & 0 & 0 & 0\\
	0 & 0 & b_N(u) & 0 & e_N(u)\kv & 0 & c_N(u) & 0 & 0\\
	\hline
	0 & d(u)\kvt1 & 0 & I_1 & 0 & 0 & 0 & 0 & 0\\
	0 & 0 & e_N(u)\kvtt & 0 & \Xm_{12}(u) & 0 & e_N(u)\kvtt & 0 & 0\\
	0 & 0 & 0 & 0 & 0 & I_1 & 0 & d(u)\kvt1 & 0\\
	\hline
	0 & 0 & c_N(u) & 0 & e_N(u)\kv & 0 & b_N(u) & 0 & 0\\
	0 & 0 & 0 & 0 & 0 & d(u)\kvt2 & 0 & I_2 & 0\\
	0 & 0 & 0 & 0 & 0 & 0 & 0 & 0 & a(u)
  \end{array}
	\right),
\end{multline*}
where
\begin{align*}
  a(u) &= 1+d(u) & b_N(u) &= 1+e_N(u) & c_N(u) &= d(u)+e_N(u).
\end{align*}
and
\begin{align*}
  \kv &= \sum_{i=3}^{2N-2}\ket{i}\otimes\ket{i}, &
  \kvt1 &= \sum_{i=3}^{2N-2}\bra{i}\otimes\ket{i}, &
  \kvt2 &= \sum_{i=3}^{2N-2}\ket{i}\otimes\bra{i}, &
  \kvtt &= \vec{k}_{12} = \sum_{i=3}^{2N-2}\bra{i}\otimes\bra{i}.
\end{align*}
Let's write the reflection and the single and the double row monodromy matrix in the following form:
\begin{align*}
	K &= \begin{pmatrix}
       Y & 0 & 0 \\
	   0 & \Ym & 0 \\
	   0 & 0 & Y^*
	  \end{pmatrix},&
	T_{AB} &= \begin{pmatrix}
		\alpha    &  \betav^t  & \beta\\
		\gammav^* &  \alpham   & \betav^*\\
		\gamma    &  \gammav^t & \alpha^*
	 \end{pmatrix}, &
	M_{AB} &= \begin{pmatrix}
       A     &  \Bv^t & B \\
	   \Cv^* &  \Am   & \Bv^* \\
	   C     &  \Cv^t & A^*
	  \end{pmatrix},
\end{align*} 
where $A$, $A^*$, $B$, $C$ are numbers $\Bv^t$, $\Cv^t$ are $1\times (2N-2)$ matrices $\Bv^*$, $\Cv^*$ are $(2N-2)\times 1$ matrices $\Am$ is a $(2N-2)\times (2N-2)$ matrix in the auxiliary space.
We will also use the following notation: $\vec{B}=\Bv^t$.

We can express the double row monodromy matrix with the single row monodromy matrix:
\[
 M_0 = T_0 K_0^R \hat{T}_0.
\]
We want to diagonalize the double row monodromy matrix:
\[
	D(\theta) = Y^{L}(\hth)A(\theta) + \mathrm{tr}_0 \Ym^{L}_0(\hth) \Am_0(\theta) + Y^{*L}(\hth)A^*(\theta). 
\]
We also know that:
\begin{align*}
	A &= \alpha Y^{R} \halpha + \betav^t \Ym^{R} \hgammav^* + \beta Y^{*R} \hgamma,\\
	\Am &= \alpham Ym^{R} \halpham + \gammav^* Y^{R} \hbetav^t + \betav^* Y^{*R} \hgammav^t,\\ 
	A^* &= \alpha^* Y^{*R} \halpha_{\begy} + \gammav^t \Ym^{R} \hbetav^* + \gamma Y^{R} \hbeta.
\end{align*}
If we want to know how these operators act on the pseudovacuum we have to 
change the order of some $\beta$ and $\gamma$ operators.
This can be done by using the Yang-Baxter relation.
The results are the following:
\begin{align*}
	A(\theta) = & \alpha(\theta) Y^{R}(\theta) \halpha(\theta) + \dots ,\\
	\Am(\theta) = & \alpham(\theta) \left[ \Ym^{R}(\theta) -
	            \frac{d(2\theta)}{a(2\theta)} Y^{R}(\theta) \right] \halpham(\theta) +
				\frac{d(2\theta)}{a(2\theta)} \alpha(\theta) Y^{R}(\theta) \halpha(\theta) \dots ,\\ 
	A^*(\theta) = & \alpha^*(\theta) \left[ Y^{*R}(\theta) +
	            \frac{d(2\hth)}{a(2\hth)} \mathrm{tr} \Ym^{R}(\theta) +
				\frac{c_N(2\hth)}{a(2\hth)} Y^{R}(\theta) \right] \halpha^*(\theta) +\\
				& + \frac{c_N(2\theta)}{a(2\theta)} \alpha(\theta) Y^{R}(\theta) \halpha(\theta) -
				\frac{d(2\hth)}{a(2\hth)} \mathrm{tr}\left\{\alpham(\theta) \left[ \Ym^{R}(\theta) - \frac{d(2\theta)}{a(2\theta)}Y^{R}(\theta) \right] \halpham(\theta) \right\} + \dots ,
\end{align*}
where the $\dots$ means terms which annihilate the pseudovacuum.
The derivation can be found in Appendix \ref{appenbc}.
It is convenient to redefine these operators:
\begin{align}
  \bar{\Am}(\theta) &= \Am(\theta) - \frac{d(2\theta)}{a(2\theta)} A(\theta) , \label{eq:defAab}\\
  \bar{A}^*(\theta) &= A^*(\theta) + \frac{d(2\hth)}{a(2\hth)}
					\mathrm{tr} \bar{\Am}(\theta) -
					\frac{c_N(2\theta)}{a(2\theta)} A(\theta) .\label{eq:defAbegy}
\end{align}
These new operators act on the pseudo-vacuum as:
\begin{align*}
	A(\theta)\ket{1}_c^{\otimes L} &= Y^R(\theta) a(\theta)^{2L}
	\ket{1}_c^{\otimes L}, \\
	\bar{\Am}(\theta)\ket{1}_c^{\otimes L} &= \left[ \Ym^{R}(\theta) -
	\frac{d(2\theta)}{a(2\theta)} Y^{R}(\theta) \right]
	\ket{1}_c^{\otimes L} \\
	\bar{A}^*(\theta)\ket{1}_c^{\otimes L} &= \left[ Y^{*R}(\theta) +
	\frac{d(2\hth)}{a(2\hth)} \mathrm{tr} \Ym^{R}(\theta) +
	\frac{c_N(2\hth)}{a(2\hth)} Y^{R}(\theta) \right]
	a(\hth)^{2L} \ket{1}_c^{\otimes L} . 
\end{align*}
Let's use these formulas in the expression of the double row transfer matrix:
\begin{align*}
	D(\theta) = & \left( Y^{L}(\hth) + \frac{d(2\theta)}{a(2\theta)} \mathrm{tr} \Ym^{L}(\hth) +
			  \frac{c_N(2\theta)}{a(2\theta)} Y^{*L}(\hth) \right) A(\theta) +\\
			  & + \mathrm{tr}\left\{\left( \Ym^{L}(\hth) - \frac{d(2\hth)}{a(2\hth)}Y^{*L}(\hth) \right) 
			  \bar{\Am}(\theta) \right\} + Y^{*L}(\hth) \bar{A}^*(\theta).
\end{align*}
At this point we know how the terms of the double row monodromy matrix act on the pseudovacuum.
The next step is to figure out the commutation relation of the elements of the transfer matrix.
This can be done by using the boundary Yang-Baxter equation.
We concentrate only on the ''wanted'' terms.
\begin{align*}
  A(\theta) \Bv^t(u) &= \frac{a(u-\theta)}{a(u+\theta)} \Bv^t(u) A(\theta) + \dots,\\
  \bar{\Am}_{1}(\theta) \Bv^t_2(u) &= \Bv^t_2(u) R_{12}^{(1)}(\theta+u-2) \bar{\Am}_{1}(\theta)
									  R_{12}^{(1)}(\theta-u) + \dots,\\
	\bar{A}^*(\theta) \Bv^t(u) &= \frac{a(u-\hth)}{a(u+\hth)} \Bv^t(u) \bar{A}^*(\theta) + \dots,
\end{align*}
where $R^{(1)}$ denotes the R-matrix of the $\On{2N-2}$ model:
\[
	R^{(1)}_{12}(u) = I_{12} - \frac{2}{u} P_{12} - \frac{2}{2N-4-u} K_{12} = 
  I_{12} + d(u) P_{12} + e_{N-1}(u) K_{12}.
\]
The derivation can be found in Appendix \ref{appenAB}.
The Bethe-states are more complicated than in the periodic case \cite{Martins:1997wb}.
To calculate these, we use the property that the vector must be symmetrical to the exchanges of the rapidities.
The derivations are found in the Appendix \ref{app:2part} and \ref{app:3part}. 
The one-particle state is simple:
\[
  \ket{\Psi}^{1-particle} = \vec{B}(\ui11)\Fvec\ket{1}_c^{\otimes L}=
  \vec{\Phi}^{(1)}(\ui11)\Fvec\ket{1}_c^{\otimes L},
\]
where $\Fvec$ is a $2N-2$ component vector.
The two particle state is the following:
\begin{align*}
	\ket{\Psi}^{2-particle} = 
	\vec{\Phi}^{(2)}_{12}(\ui11,\ui12)\Fvec_{12}\ket{1}_c^{\otimes L},
\end{align*}
where
\begin{multline*}
	\vec{\Phi}^{(2)}_{12}(\ui11,\ui12) =\\
   \vec{B}_1(\ui11)\vec{B}_2(\ui12) -
  \frac{1}{a(2\ui12)} \frac{e_N(\ui1{12})}{b_N(\ui1{12})} B(\ui11)\vec{k}_{12} A(\ui12) +
  e_N(\wi1{12})B(\ui11) \vec{k}_{12} \bar{\Am}_{2}(\ui12),
\end{multline*}
and $\ui1{12}=\ui11-\ui12$ and $\wi1{12}=\ui11+\ui12$.
This is not analogous to the periodical case.
The third term which contains $\Am$ is present only in  the model with a boundary.

The $n_1$-particle state is the following:
\begin{align*} 
  &\vec{\Phi}^{(n_1)}_{1\dots n_1}(\ui11,\dots,\ui1{n_1}) =
  \vec{B}_1(\ui11)\vec{\Phi}^{(n_1-1)}_{2\dots n_1}(\ui12,\dots,\ui1{n_1}) -\\
   & - \sum_{j=2}^{n_1} \frac{1}{a(2\ui1j)} \frac{e_N(\ui1{1j})}{b_N(\ui1{1j})}
   \left[ \prod_{\substack{k=2\\k\neq j}}^{n_1} \frac{a(\ui1{kj})}{a(\wi1{kj})} \right] 
   \vec{k}_{1j}B(\ui11) \times \\
   & \times 
  \vec{\Phi}^{({n_1}-2)}_{2,\dots,\hat{\jmath},\dots,{n_1}}(\ui12,\dots,\widehat{\ui1{j}},\dots,\ui1{n_1})A(u_j)
  \prod_{k=2}^{\substack{j-1\\ \longrightarrow}} \frac{R_{kj}^{(1)}(\ui1{kj})}{a(\ui1{kj})} +\\
   & + \sum_{j=2}^n e_N(\wi1{1j}) 
  \left[\prod_{\substack{k=2\\k\neq j}}^{n_1} a(\ui1{kj}) \right] \vec{k}_{1j}B(\ui11) \times \\
   & \times \vec{\Phi}^{({n_1}-2)}_{2,\dots,\hat{\jmath},\dots,{n_1}}(\ui12,\dots,\widehat{\ui1{j}},\dots,\ui1{n_1})
  \prod_{\substack{k=2\\k\neq j}}^{\substack{{n_1}\\ \longrightarrow}} R_{kj}^{(1)}(\wi1{kj}-2) \bar{\Am}_{j}(\ui1j) 
  \prod_{k=j+1}^{\substack{{n_1}\\ \longleftarrow}} \frac{R_{jk}^{(1)}(\ui1{jk})}{a(\ui1{jk})}, 
\end{align*}
where hats denote the missing terms.

The eigenvalue of the transfer matrix can be written in the following form:
\[
	\lambda(\theta) = F_0(\theta) + \lambda_1(\theta) + \bar{F}_0(\theta), 
\]
where 
\begin{align*}
	F_0(\theta) &= 
	k_0^{L}(\hth)k_0^{R}(\theta) a(\theta)^{2L} 
	\prod_{i=1}^{n_1} \frac{a(\ui1{i}-\theta)}{a(\ui1{i}+\theta)},\\
	\bar{F}_0(\theta) &= \bar{k}_0^{L}(\hth)\bar{k}_0^{R}(\theta)  
	a(\hth)^{2L} \prod_{i=1}^{n_1} 
	\frac{a(\ui1{i}-\hth)}{a(\ui1{i}+\hth)},
\end{align*}
and
\begin{align*}
	k_0^{L}(\theta) &= Y^{L}(\theta) + \frac{d(2\hth)}{a(2\hth)} \mathrm{tr } \Ym^{L}(\theta) +
					 \frac{c_N(2\hth)}{a(2\hth)} Y^{*L}(\theta) ,\\
	\bar{k}_0^{L}(\theta) &= Y^{*L}(\theta) ,\\
	k_0^{R}(\theta) &= Y^{R}(\theta) ,\\
	\bar{k}_0^{R}(\theta) &= Y^{*R}(\theta) +
						   \frac{d(2\hth)}{a(2\hth)} \mathrm{tr} \Ym^{R}(\theta) +
						   \frac{c_N(2\hth)}{a(2\hth)} Y^{R}(\theta) .
 \end{align*}
and the $\lambda_1$ is the eigenvalue of a new transfer matrix. This is a transfer matrix of the $\On{2N-2}$ model with shifted rapidities and new reflection matrix:
\begin{align*}
	K^{(1)L}(\theta-1) &= \Ym^{L}(\theta) -
						\frac{d(2\theta)}{a(2\theta)} Y^{*L}(\theta),\\
	K^{(1)R}(\theta-1) &= \Ym^{R}(\theta) -
						\frac{d(2\theta)}{a(2\theta)} Y^{R}(\theta).
\end{align*}

\subsubsection{The first step}
The new transfer matrix:
\[
  D^{(1)}(\theta) = \mathrm{tr}_0 \left[ K^{(1)L}_0(2N-3-\theta) \prod_{i=1}^{n_1} R^{(1)}_{0i}(\theta-\ui1{i}) K^{(1)R}_0(\theta-1) \prod_{i=n_1}^{1} R^{(1)}_{i0}(\theta+\ui1{i}-2) \right],
\]
where $n_1$ is the number of magnons at the first step of the nesting.
A new basis should also be introduced at this step: $\{\ket{3},\ket{4},\ket{5},\dots,\ket{2N}\}\longrightarrow\{\ket{2}_c,\ket{5},\dots,\ket{2N},\ket{\bar{3}}_c\}$ where
\begin{align*}
	\frac{1}{\sqrt{2}}\left(\ket{3}+i\ket{4}\right) \longrightarrow \ket{2}_c,\\
	\frac{1}{\sqrt{2}}\left(\ket{3}-i\ket{4}\right) \longrightarrow \ket{\bar{2}}_c.
\end{align*}
We have to decompose the K-matrix
\begin{align*}
  K^{(1)} = \begin{pmatrix}
	   Y^{(1)} & 0 & 0 \\
	   0 & \Ym^{(1)} & 0 \\
	   0 & 0 & Y^{*(1)}
	  \end{pmatrix}.
\end{align*} 
The eigenvalue of this transfer matrix (similar to $D$) is:
\[
	\lambda_1(\theta) = F_1(\theta) + \lambda_2(\theta) + \bar{F}_1(\theta), 
\]
where
\begin{align*}
	F_1(\theta) &= k_1^{L}(\hth) k_1^{R}(\theta) \prod_{i=1}^{n_1} 
	a(\theta-\ui1{i})a(\theta+\ui1{i}-2) \prod_{i=1}^{n_2} 
	\frac{a(\ui2{i}-\theta)}{a(\ui2{i}+\theta-2)},\\
	\bar{F}_1(\theta) &= \bar{k}_1^{L}(\hth) \bar{k}_1^{R}(\theta) 
	\prod_{i=1}^{n_1} a(\hth-\ui1{i})a(\hth+\ui1{i}-2) \prod_{i=1}^{n_2} 
	\frac{a(\ui2{i}-\hth)}{a(\ui2{i}+\hth-2)},
\end{align*}
and
\begin{align*}
	k_1^{L}(\theta) &= Y^{(1)L}(\theta-1) +
					 \frac{d(2\hth-2)}{a(2\hth-2)} \mathrm{tr} \Ym^{(1)L}(\theta-1) +
					 \frac{c_{N-1}(2\hth-2)}{a(2\hth-2)} Y^{*(1)L}(\theta-1) ,\\
	\bar{k}_1^{L}(\theta) &= Y^{*(1)L}(\theta-1) ,\\
	k_1^{R}(\theta) &= Y^{(1)R}(\theta-1) ,\\
	\bar{k}_1^{R}(\theta) &= Y^{*(1)R}(\theta-1) +
						   \frac{d(2\hth-2)}{a(2\hth-2)} \mathrm{tr} \Ym^{(1)R}(\theta-1) +
						   \frac{c_{N-1}(2\hth-2)}{a(2\hth-2)} Y^{(1)R}(\theta-1) .
 \end{align*}

\subsubsection{The $k$th step}
 This procedure can be continued until the O(4) model is reached.
 The K-matrix at the $k$th step is:
\begin{align*}
	K^{(k)L}(\theta-k) &= \Ym^{(k-1)L}(\theta-(k-1)) -
						\frac{d(2\theta-2(k-1))}{a(2\theta-2(k-1))} Y^{*(k-1)L}(\theta-(k-1)),\\
	K^{(k)R}(\theta-k) &= \Ym^{(k-1)R}(\theta-(k-1)) -
						\frac{d(2\theta-2(k-1))}{a(2\theta-2(k-1))} Y^{(k-1)R}(\theta-(k-1)).
\end{align*}
The new basis at the $k$th step is the following: $\{\ket{2k+1},\ket{2k+2},\ket{2k+3},\dots,\ket{2N}\}\longrightarrow\{\ket{k+1}_c,\ket{2k+3},\dots,\ket{2N},\ket{\overline{k+1}}_c\}$ where
\begin{align*}
	\frac{1}{\sqrt{2}}\left(\ket{2k+1}+i\ket{2k+2}\right) \longrightarrow \ket{k+1}_c,\\
	\frac{1}{\sqrt{2}}\left(\ket{2k+1}-i\ket{2k+2}\right) \longrightarrow \ket{\overline{k+1}}_c.
\end{align*}
The components of the K-matrix are
\begin{align*}
  K^{(k)} = \begin{pmatrix}
	   Y^{(k)} & 0 & 0 \\
	   0 & \Ym^{(k)} & 0 \\
	   0 & 0 & Y^{*(k)}
	  \end{pmatrix}.
\end{align*} 
\[
  D^{(k)}(\theta) = \mathrm{tr}_0 \left[ K^{(k)L}_0(2N-2-k-\theta) \prod_{i=1}^{n_k} R^{(k)}_{0i}(\theta-\ui{k}{i}) K^{(k)R}_0(\theta-k) \prod_{i=n_k}^{1} R^{(k)}_{i0}(\theta+\ui{k}{i}-2k) \right],
\]
\[
	\lambda_k(\theta) = F_k(\theta) + \lambda_{k+1}(\theta) + \bar{F}_k(\theta), 
\]
where
\begin{align*}
	F_k(\theta) &= k_k^{L}(\hth) k_k^{R}(\theta) \prod_{i=k}^{n_k} 
	a(\theta-\ui{k}{i}) a(\theta+\ui{k}{i}-2k) \prod_{i=1}^{n_{k+1}} 
	\frac{a(\ui{k+1}{i}-\theta)}{a(\ui{k+1}{i}+\theta-2k)},\\
	\bar{F}_k(\theta) &= \bar{k}_k^{L}(\hth) \bar{k}_k^{R}(\theta) 
	\prod_{i=k}^{n_k} a(\hth-\ui{k}{i})a(\hth+\ui{k}{i}-2k) 
	\prod_{i=1}^{n_{k+1}} \frac{a(\ui{k+1}{i}-\hth)}{a(\ui{k+1}{i}+\hth-2k)},
\end{align*}
and
\begin{align*}
	k_k^{L}(\theta) &= Y^{(k)L}(\theta-k) +
					\frac{d(2\hth-2k)}{a(2\hth-2k)} \mathrm{tr} \Ym^{(k)L}(\theta-k) +
					\frac{c_{N-k}(2\hth-2k)}{a(2\hth-2k)} Y^{*(k)L}(\theta-k) ,\\
	\bar{k}_k^{L}(\theta) &= Y^{*(k)L}(\theta-k) ,\\
	k_k^{R}(\theta) &= Y^{(k)R}(\theta-k) ,\\
	\bar{k}_k^{R}(\theta) &= Y^{*(k)R}(\theta-k) +
						  \frac{d(2\hth-2k)}{a(2\hth-2k)} \mathrm{tr} \Ym^{(k)R}(\theta-k) +
						  \frac{c_{N-k}(2\hth-2k)}{a(2\hth-2k)} Y^{(k)R}(\theta-k) .
 \end{align*}
where the $k<N-2$ and the $\lambda_{N-2}$ is the eigenvalue of the transfer matrix of the O(4) model.

\subsubsection{The $(N-2)$th step}
At the last step of the nesting we have an $\On4$ symmetric model which can be easily solved if we use the $\mathfrak{su}(2)\otimes\mathfrak{su}(2)$ basis.
The factorized R-matrix is:
\[
	R^{(N-2)}(\theta) = \frac{1}{a(\theta)} \left[ \mathbb{I} +d(\theta) \mathbb{P} \right] \otimes \left[ \mathbb{I} + d(\theta) \mathbb{P} \right] =
	\frac{1}{a(\theta)} R^{XXX}(\theta) \otimes R^{XXX}(\theta).
\]
For the \ref{Dn_a}, \ref{Dn_b} and \ref{Dn_c} reflections the K-matrices are 
factorized too
\[
	K^{(N-2)}(\theta) = K_0(\theta)
  \begin{pmatrix}
	K_A^+(\theta) & 0\\
	0 & K_D^+(\theta)
	\end{pmatrix}
	\otimes
  \begin{pmatrix}
	K_A^-(\theta) & 0\\
	0 & K_D^-(\theta)
	\end{pmatrix} = K_0(\theta)
	K^+(\theta) \otimes K^-(\theta).
\]
The transfer matrix can be written in the following form:
\[
	D^{(N-2)}(\theta)= K_0^L(N-\theta)K_0^R(\theta-N+2) \prod_{i=1}^{n_{N-2}} \frac{1}{a(\theta-\ui{N-2}{i})} \frac{1}{a(\theta+\ui{N-2}{i}-2N+4)} D_+(\theta)D_-(\theta),
\]
where $D_+$ and $D_-$ are double row transfer matrices of XXX spin chains:
\begin{multline*}
  D_\pm(\theta) =\\
   \mathrm{tr}_0 \left[ K^{L,\pm}_0(N-\theta) \prod_{i=1}^{n_k} R^{XXX}_{0i}(\theta-\ui{N-2}{i}) K^{R,\pm}_0(\theta-N+2) \prod_{i=n_k}^{1} R^{XXX}_{i0}(\theta+\ui{N-2}{i}-2N+4) \right],
\end{multline*}
The eigenvalue of the $\On4$ transfer matrix ($\lambda_{N-2}$):
\[
  \lambda_{N-2}(\theta) = K_0^L(N-\theta)K_0^R(\theta-N+2) \prod_{i=1}^{n_{N-2}} \frac{\theta-\ui{N-2}{i}}{\theta-\ui{N-2}{i}-2} \frac{\theta+\ui{N-2}{i}-2N+4}{\theta+\ui{N-2}{i}-2N+2} \lambda_+(\theta) \lambda_-(\theta),
\]
where
\begin{multline*}
  \lambda_\pm(\theta) = k_A^{L,\pm} k_A^{R,\pm} 
		\prod_{i=1}^{n_{N-2}}	\frac{\theta-\ui{N-2}{i}-2}{\theta-\ui{N-2}{i}}
								\frac{\theta+\ui{N-2}{i}-2N+2}{\theta+\ui{N-2}{i}-2N+4} \times\\
  \times \prod_{i=1}^{n_\pm}	\frac{\theta-\ui{\pm}{i}+2}{\theta-\ui{\pm}{i}}
								\frac{\theta+\ui{\pm}{i}-2N+4}{\theta+\ui{\pm}{i}-2N+2} +\\
	+ k_D^{L,\pm} k_D^{R,\pm}	\frac{\theta-N+2}{\theta-N+1}
		\prod_{i=1}^{n_\pm}		\frac{\theta-\ui{\pm}{i}-2}{\theta-\ui{\pm}{i}}
								\frac{\theta+\ui{\pm}{i}-2N}{\theta+\ui{\pm}{i}-2N+2}.
\end{multline*}
and
\begin{align*}
	k_A^{L,\pm} &= K_A^{L,\pm}(N-\theta) + \frac{d(2\theta-2N+4)}{a(2\theta-2N+4)} K_D^{L,\pm}(N-\theta), \\
	k_A^{R,\pm} &= K_A^{R,\pm}(N-\theta), \\
	k_D^{L,\pm} &= K_D^{L,\pm}(\theta-N+2), \\
	k_D^{R,\pm} &= K_D^{R,\pm}(\theta-N+2) - \frac{d(2\theta-2N+4)}{a(2\theta-2N+4)} K_A^{R,\pm}(\theta-N+2). \\
\end{align*}

\subsubsection{$\On{2M}\times\On{2N-2M}$ symmetric boundaries } \label{subsec:ABA}
In this subsection we use the following K-matrix:
\[
 K(u) = \mathrm{diag}( \underbrace{c(u),\dots,c(u)}_{2M}, \underbrace{1,\dots,1}_{2N-2M} ),
\]
where
\[
 c(u) = \frac{N-2M+u}{N-2M-u}.
\]
The K-matrix in the $k$th nesting step for $k<M$ is the following:
\[
 K^{(k)}(u) =\frac{u+k}{u} \frac{N-2M+k-u}{N-2M-k-u} \mathrm{diag}( \underbrace{c_k(u),\dots,c_k(u)}_{2M-2k}, \underbrace{1,\dots,1}_{2N-2M} ),
\]
where
\[
 c_k(u) = \frac{N-2M+k+u}{N-2M+k-u}.
\]
if $k\ge M$ then
\[
 K^{(k)}(u) =\frac{u+k}{u} \frac{N-k-u}{N-2M-k-u} \mathrm{diag}(\underbrace{1,\dots,1}_{2N-2k}) .
\]
The explicit form of the boundary coefficients for $k< M$ are:
\begin{align*}
  k_k^{L}(\theta) =& \bar{k}_k^{R}(\theta) = Y^{(k)}(\theta-k) +
				  \frac{d(2\hth-2k)}{a(2\hth-2k)} \mathrm{tr} \Ym^{(k)}(\theta-k) +
				  \frac{c(2\hth-2k)}{a(2\hth-2k)} Y^{(k)}(\theta) = \\
				=& -\frac{(\hth-N+2)(\hth+N-2M)(\hth-2N+2)}{(\hth-k-1)(\hth-N+1)(\hth-N-2M+2)},\\
  \bar{k}_k^{L}(\theta) =& k_k^{R}(\theta) = Y^{(k)}(\theta) = - \frac{\theta(\theta+N-2M)}{(\theta-k)(\theta-N+2M)},
 \end{align*}
 while for $k\ge M$ are
\begin{align*}
  k_k^{L}(\theta) =& \bar{k}_k^{R}(\theta) = \frac{(\hth-N+2)(\hth-N)(\hth-2N+2)}{(\hth-k-1)(\hth-N+1)(\hth-N-2M+2)},\\
  \bar{k}_k^{L}(\theta) =& k_k^{R}(\theta) = \frac{\theta(\theta-N)}{(\theta-k)(\theta-N+2M)},
 \end{align*}
We still need the coefficients of the last step.
If $M<N-1$
\begin{align*}
	K_0(\theta) &= \frac{\theta+N-2}{\theta}\frac{\theta-2}{\theta+2M-2},\\
	K_A^\pm(\theta) &= 1, &	K_D^\pm(\theta) &= 1,\\
	k_A^\pm			&= 1, &	k_D^\pm			&= 1,\\
	\bar{k}_A^{\pm} &= \frac{N-\theta}{N-1-\theta}, &
	\bar{k}_D^{\pm} &= \frac{\theta-N+2}{\theta-N+1},
\end{align*}
and if $M=N-1$
\begin{align*}
	K_0(\theta) &= \frac{\theta+N-2}{\theta+2N-4},\\
	K_A^\pm(\theta) &= -1, & K_D^\pm(\theta) &= 1,\\
	k_A^\pm         &= -1, & k_D^\pm         &= 1,\\
	\bar{k}_A^{\pm} &= -\frac{\theta-N+2}{\theta-N+1}, &
	\bar{k}_D^{\pm} &= \frac{N-\theta}{N-1-\theta},
\end{align*}
The BAEs can be easily calculated using the vanishing residue condition:
\begin{align} \label{eq:res}
 \operatorname*{Res}_{\theta=\ui{k}{i}} \lambda(\theta) = 0.
 \end{align}
It is convenient to redefine the rapidity variables:
\begin{align*}
 \vi{k}{i} &= \ui{k}{i} -k, \text{ where } k=1,\dots,N-2, &
 \vi{\pm}{i} &= \ui{\pm}{i} -N-1. 
\end{align*}
The BAEs are the following:
\begin{align*}
	r_1(\vi{1}{j})^2 \left( \frac{\vi{1}{j}-1}{\vi{1}{j}+1} \right)^{2L} 
	\prod_{k} \prod_{\substack{i=1\\(i,k)\neq(j,1)}}^{n_k}   
	s_{k1}(\vi{1}{j}-\vi{k}{i}) s_{k1}(\vi{1}{j}+\vi{k}{i}) = 1
\end{align*}
for $l=1$ and
\begin{align*}
	r_l(\vi{l}{j})^2 \prod_{k} \prod_{\substack{i=1\\(i,k)\neq(j,l)}}^{n_k}   
	s_{kl}(\vi{l}{j}-\vi{k}{i}) s_{kl}(\vi{l}{j}+\vi{k}{i}) = 1
\end{align*}
for $l=2,\dots,N-2,+,-$
where $k=1,\dots,N-2,+,-$.
This result is the same as in a previous publication \cite{Gombor:2015kdu}.

\subsubsection{$\mathrm{U}(N)$ symmetric boundaries } \label{subsec:UN}
In this subsection we use the following K-matrix:
\[
  K(u) = 
  \begin{pmatrix}
	c  & iu &         &     &    \\
	-iu &  c &        &     &    \\
	    &    & \ddots &     &    \\
	    &    &        &  c  & iu \\
	    &    &        & -iu &  c
  \end{pmatrix},
\]
where $K(u)$ is a $2N\times2N$ dimensional matrix.
This matrix in the complex basis looks like this:
\[
	K^{(c)}(u) = 
	\begin{pmatrix}
		(c-u)\mathbb{I}_N &                   \\
						  & (c+u)\mathbb{I}_N
	\end{pmatrix}.
\]
In this basis the residual symmetry subgroup $\mathrm{U}(N)$ of 
$\mathrm{SO}(2N)$ looks like this:
\[
	\begin{pmatrix}
		g &        \\
	      & \bar{g}
	\end{pmatrix}.
\]
where $\bar{g}$ is the complex conjugate of $g$ and $g$ is in the defining 
representation of the $\mathrm{U}(N)$ group.

The K-matrices in the transfer matrix are the following: $K^R=(K^L)^T=K$ (in 
the real basis) and consequently $Y^R=Y^{*L}$ and $Y^L=Y^{*R}$.
For this choice the following equations exist in any step of the nesting: $K^{(k)R}=(K^{(k)L})^T=K^{(k)}$, $Y^{(k)R}=Y^{*(k)L}$ and $Y^{(k)L}=Y^{*(k)R}$.
At the $k$th nesting step the K-matrix is the following:
\[
 K^{(k)}(u) =\frac{u+k}{u} 
  \begin{pmatrix}
	c-k & iu  &        &     &    \\
	-iu & c-k &        &     &    \\
	    &     & \ddots &     &    \\
	    &     &        & c-k & iu \\
	    &     &        & -iu & c-k
  \end{pmatrix},
\]
where $K^{(k)}(u)$ is a $2N-2k\times2N-2k$ dimensional matrix.
The explicit form of the boundary coefficients:
\begin{align*}
  k_k^{L}(\theta) =& \bar{k}_k^{R}(\theta) = \frac{(c-\hth)(\hth-N)(\hth-2N+2)}{(\hth-k-1)(\hth-N+1)},\\
  \bar{k}_k^{L}(\theta) =& k_k^{R}(\theta) = \frac{\theta(c-\theta)}{\theta-k},
 \end{align*}
\begin{align*}
	K_0(\theta)   &= -\frac{\theta+n-2}{\theta}  ,\\
	K_A^-(\theta) &= 1 & K_A^+(\theta) &= N-2-c+\theta ,\\
	K_D^-(\theta) &= 1 & K_D^+(\theta) &= N-2-c-\theta ,\\
	k_A^{R,-}     &= 1 & k_A^{R,+}(\theta) &= -c+\theta ,\\
	k_D^{L,-}     &= 1 & k_D^{L,+}(\theta) &= 2N-2-c-\theta\ ,\\
	k_A^{L,-} &= \frac{N-\theta}{N-1-\theta} & k_A^{L,+} &= \frac{N-\theta}{N-1-\theta}(-c+\theta),\\
	k_D^{R,-} &= \frac{\theta-N+2}{\theta-N+1} & k_D^{R,+} &= \frac{\theta-N+2}{\theta-N+1}(2N-2-c-\theta),
\end{align*}
Similarly to the previous case the Bethe Ansatz equations can be calculated from \eqref{eq:res}.

\subsubsection{$\On{2}\times\On{2N-2}$ symmetric boundaries } \label{subsec:u1}
In this subsection we use the following K-matrix:
\[
  K(u) = 
  \begin{pmatrix}
	\frac{(N-2)^2-c^2-u^2}{(N-2-u)^2-c^2} & i\frac{2cu}{(N-2-u)^2-c^2} & & & \\
	-i\frac{2cu}{(N-2-u)^2-c^2} &  \frac{(N-2)^2-c^2-u^2}{(N-2-u)^2-c^2} & & & \\
		&    & 1 & & \\
		&    & & \ddots & \\
		&    & &  & 1 \\
  \end{pmatrix},
\]
where $K(u)$ is a $2N\times2N$ dimensional matrix.
This matrix in the complex basis looks like this:
\[
	K^{(c)}(u) = 
	\begin{pmatrix}
		\frac{N-2+c+u}{N-2+c-u} &   &        &   &                         \\
		                        & 1 &        &   &                         \\
		                        &   & \ddots &   &                         \\	
		                        &   &        & 1 &                         \\	
						        &   &        &   & \frac{N-2-c+u}{N-2-c-u}
	\end{pmatrix}.
\]

Now we select the matrices as in the previous case: $K^R=(K^L)^T=K$.
At the $k$th nesting step the K-matrix is the following:
\[
 K^{(k)}(u) =\frac{u+k}{u} \frac{N-k-c-u}{N-2-k-c-u} \mathrm{diag}(\underbrace{1,\dots,1}_{2N-2k}) .
\]
where $K^{(k)}(u)$ is $2N-2k\times2N-2k$ dimensional matrix.
The explicit form of the boundary coefficients for $k=0$:
\begin{align*}
  k_0^{L}(\theta) =& \bar{k}_0^{R}(\theta) = -\frac{(\hth-N+2-c)(\hth+N-2-c)(\hth-2N+2)(\hth-n)}{(\hth-1)(\hth-N+1)(\hth-N+c)(\hth-N-c)},\\
  \bar{k}_0^{L}(\theta) =& k_0^{R}(\theta) = - \frac{\theta+N-2-c}{\theta-N+2+c},
 \end{align*}
and if $k>0$:
\begin{align*}
  k_k^{L}(\theta) =& \bar{k}_k^{R}(\theta) = \frac{(\hth-N+2-c)(\hth-N)(\hth-2N+2)}{(\hth-k-1)(\hth-N+1)(\hth-N-c)},\\
  \bar{k}_k^{L}(\theta) =& k_k^{R}(\theta) = \frac{\theta(\theta-N+c)}{(\theta-k)(\theta-N+2+c)},
 \end{align*}
 and
\begin{align*}
	K_0(\theta) &= \frac{\theta+N-2}{\theta}\frac{\theta-2+c}{\theta+c},\\
	K_A^\pm(\theta) &= 1, & K_D^\pm(\theta) &= 1,\\
	k_A^\pm         &= 1, &	k_D^\pm			&= 1,\\
	\bar{k}_A^{\pm} &= \frac{N-\theta}{N-1-\theta}, &
	\bar{k}_D^{\pm} &= \frac{\theta-N+2}{\theta-N+1},
\end{align*}
Similarly to the previous case the Bethe Ansatz equations can be calculated from \eqref{eq:res}.

\subsection{Nesting for  $\On{2M+1}\times\On{2N-2M-1}$ symmetric boundaries} 
\label{sub4}
In this subsection we use the following K-matrix:
\[
 K(u) = \mathrm{diag}( \underbrace{c(u),\dots,c(u)}_{2M}, \underbrace{1,\dots,1}_{2N-2M-1},c(u) ),
\]
where
\[
 c(u) = \frac{N-2M-1+u}{N-2M-1-u}.
\]
At the $k$th nesting step if $k<M$ the K-matrix is the following:
\[
 K^{(k)}(u) =\frac{u+k}{u} \frac{N-2M-1+k-u}{N-2M-1-k-u} \mathrm{diag}( \underbrace{c_k(u),\dots,c_k(u)}_{2M-2k}, \underbrace{1,\dots,1}_{2N-2M-1},c_k(u) ),
\]
where
\[
 c_k(u) = \frac{N-2M-1+k+u}{N-2M-1+k-u}.
\]
if $k\ge M$ than
\[
 K^{(k)}(u) =\frac{u+k}{u} \frac{N-1-k-u}{N-2M-1-k-u} \mathrm{diag}(\underbrace{1,\dots,1}_{2N-2k-1},\frac{N-1-k+u}{N-1-k-u}) .
\]
The explicit form of the boundary coefficients for $k< M$:
\begin{align*}
  k_k^{L}(\theta) =& \bar{k}_k^{R}(\theta) = -\frac{(\hth-N+2)(\hth+N-2M-1)(\hth-2N+2)}{(\hth-k-1)(\hth-N+1)(\hth-N-2M+1)},\\
  \bar{k}_k^{L}(\theta) =& k_k^{R}(\theta) = - \frac{\theta(\theta+N-2M-1)}{(\theta-k)(\theta-N+2M+1)},
 \end{align*}
 and if $k\ge M$
\begin{align*}
  k_k^{L}(\theta) =& \bar{k}_k^{R}(\theta) = \frac{(\hth-N+2)(\hth-2N+2)}{(\hth-k-1)(\hth-N-2M+1)},\\
  \bar{k}_k^{L}(\theta) =& k_k^{R}(\theta) = \frac{\theta(\theta-N+1)}{(\theta-k)(\theta-N+2M+1)},
 \end{align*}
At the last step of the nesting we have a $\On4$ symmetric model with the following boundary condition:
\begin{multline*}
	K^{(N-2)}(\theta) = \frac{\theta+N-2}{\theta+2M-1}
	\mathrm{diag}(	\frac{1-\theta}{\theta},\frac{1-\theta}{\theta},
					\frac{1-\theta}{\theta},\frac{1+\theta}{\theta}) = \\
	K_0(\theta) \mathrm{diag}(	\frac{1-\theta}{\theta},\frac{1-\theta}{\theta},
								\frac{1-\theta}{\theta},\frac{1+\theta}{\theta}).
\end{multline*}
This case differs from the foregoing that the reflection matrix does not factorize in the $\mathrm{SU}(2)\times\mathrm{SU}(2)$ basis.
The solution of this model is derived in the appendix \ref{o4sol}. The eigenvalue of this transfer matrix is the following:
\[
 \lambda_{N-2}(\theta) = K_0(N-\theta)K_0(\theta-N+2) \lambda'(\theta) \lambda'(2N-2-\theta),
\]
where
\begin{align*}
 \lambda'(\theta) =  \prod_{i=1}^{n_{N-2}} \frac{\theta-\ui{N-2}{i}-2}{\theta-\ui{N-2}{i}} \frac{\theta+\ui{N-2}{i}-2N+2}{\theta+\ui{N-2}{i}-2N+4} \prod_{i=1}^{n_{N-1}} \frac{\theta-\ui{N-1}{i}+2}{\theta-\ui{N-1}{i}} 
 + \prod_{i=1}^{n_{N-1}} \frac{\theta-\ui{N-1}{i}-2}{\theta-\ui{N-1}{i}}.
\end{align*}
In this case it is also convenient to redefine the rapidities:
\begin{align*}
 \vi{k}{i} &= \ui{k}{i} -k, & \text{ where } k&=1,\dots,N-1, 
\end{align*}
The BAEs are the following:
\begin{align*}
	r_1(\vi{1}{j})^2 \left( \frac{\vi{1}{j}-1}{\vi{1}{j}+1} \right)^{2L} 
	\prod_{k} \prod_{\substack{i=1\\(i,k)\neq(j,1)}}^{n_k}   
	s_{k1}(\vi{1}{j}-\vi{k}{i}) s_{k1}(\vi{1}{j}+\vi{k}{i}) = 1
\end{align*}
for $l=1$ and
\begin{align*}
	r_l(\vi{l}{j})^2 \prod_{k} \prod_{\substack{i=1\\(i,k)\neq(j,l)}}^{n_k}   
	s_{kl}(\vi{l}{j}-\vi{k}{i}) s_{kl}(\vi{l}{j}+\vi{k}{i}) = 1
\end{align*}
for $l=2,\dots,N-2$ and
\begin{multline*}
 \prod_{i\neq j}^{n_{N-1}}   s_{N-1,N-1}(\vi{N-1}{j}-\vi{N-1}{i}) \times \\
 \times {\prod_{i=1}^{n_{N-2}} }	s_{N-1,N-2}(\vi{N-1}{j}-\vi{N-2}{i})
									s_{N-1,N-2}(\vi{N-1}{j}+\vi{N-2}{i}) = 1, 
\end{multline*}
where the $\alpha_i$s in the scattering phase are the $\mathfrak{su}(N)$ simple roots.
\[
	\alpha_i \cdot \alpha_j = 
	\begin{pmatrix}
		 2 & -1 &    &        &    &    & \\
		-1 &  2 & -1 &        &    &    & \\
		   & -1 &  2 &        &    &    & \\
		   &    &    & \ddots &    &    & \\
		   &    &    &        &  2 & -1 & \\
		   &    &    &        & -1 &  2 & -1  \\
		   &    &    &        &    & -1 &  2 \\
	\end{pmatrix}
\]
Only one reflection factor is non-trivial:
\[
 r_M(u) = \frac{N-M-1+u}{N-M-1-u}.
\]

\section{SU(4) spin chains and the connection between the S-reflections}

Since the SU(4) group locally isomorphic to the SO(6) group, the SU(4) spin chain in the six dimensional representation is equivalent to the fundamental representation of the SO(6) spin chain.

For the SU(4) spin chain we have two S-reflections: the SO(4) and the Sp(4) symmetric reflections.
For the SO(6) spin chain there are also two S-reflections: the SO(3)$\times$SO(3) and the SO(5) symmetric reflections.
Because of $\mathfrak{so}(4)\cong\mathfrak{so}(3)\oplus\mathfrak{so}(3)$ and $\mathfrak{sp}(4)\cong\mathfrak{so}(5)$ we can see that the \ref{An_b} and the \ref{An_c} reflections are equivalent to the \ref{Dn_d} reflections for the SU(4) spin chain.

The defining representation of the R-matrix of the SU(4) spin chain is the following:
\[
	R_{\alpha\beta}(u) = I_{\alpha\beta} - \frac{2}{u} P_{\alpha\beta},
\]
where the Greek indices denote the fundamental representation.
This R-matrix has SU(4) symmetry:
\[
	\left( g \otimes g \right) R_{\alpha\beta}(u) \left( g^\dagger \otimes g^\dagger \right) = R_{\alpha\beta}(u), 
\]
where $g\in\mathrm{SU}(4)$.
We will need the R-matrix that describes the scattering between the fundamental and anti-fundamental representations.
\[
	R_{\alpha\bar{\beta}}(u) = R_{\alpha\beta}^{t_\beta}(4-u),
\]
where Greek indices with overline denote the anti-fundamental representation and $t_\beta$ means transposition with respect the $\beta$ vector space.
This R-matrix also has SU(4) symmetry:
\[
	\left( g \otimes \bar{g} \right) R_{\alpha\bar{\beta}}(u) \left( g^\dagger \otimes \bar{g}^\dagger \right) = R_{\alpha\bar{\beta}}(u), 
\]
where $\bar{g}$ is the complex conjugate of $g$.
In this section we deal with K-matrices which intertwine between the fundamental and anti-fundamental representations.
So they transform under the SU(4) transformation in the following way:
\begin{align} \label{eq:bndSym}
	K(u) \longrightarrow g K(u) \bar{g}^\dagger = g K(u) g^t.
\end{align}
The BYBE for these K-matrices is the following:
\begin{multline*}
	R_{\alpha\beta}(u_1-u_2) (K(u_1) \otimes \mathbb{I}) R_{\beta\bar{\alpha}}(u_1+u_2)
							 (\mathbb{I} \otimes K(u_2)) = \\ 
	(\mathbb{I} \otimes K(u_2))	R_{\alpha\bar{\beta}}(u_1+u_2) (K(u_1) \otimes \mathbb{I})
								R_{\beta\alpha}(u_1-u_2),
\end{multline*}
where we used the $R_{\bar{\alpha}\bar{\beta}}=R_{\alpha\beta}$ identity.
There are two solutions of this equation:
\begin{align*}
	K_1(u) &= \mathbb{I}_4, &
	K_2(u) &= \begin{pmatrix}
				0 & \mathbb{I}_2\\
				-\mathbb{I}_2 & 0								
			\end{pmatrix}.
\end{align*}
The first one is the \ref{An_b} and the second one is the \ref{An_c} case.
From equation \eqref{eq:bndSym} we can see that the residual symmetries are the SO(4) and Sp(4).

We will now construct these K-matrices in the six-dimensional representation.
The six dimensional representation can be projected out from the tensor product of two fundamental one.
We will denote by $A$ the linear transformation which  intertwines between the six dimensional representation and the tensor product of two fundamental representations that is 
$A: \mathbb{C}^4 \otimes \mathbb{C}^4 \rightarrow \mathbb{C}^6$.
We can get a six dimensional representation $G$ of $g$ using $A$: $G=A(g \otimes g)A^\dagger$.
This representation is pseudo-real which means that there is a unitary matrix $C$ such that $\bar{G}=CGC^\dagger$.
But we know that this representation is not just pseudo-real but also real. This means that there is certainly a matrix $B$ such that $G_r=BGB^\dagger$ where $\bar{G}_r=G_r$.

Then the six dimensional representations of the K-matrices can be calculated:
\[
	K^{(6)}(u) = BA(\mathbb{I} \otimes K(u-1))R_{\alpha\beta}(2u)(K(u+1) \otimes \mathbb{I})P A^\dagger C^\dagger B^\dagger.
\]
This formula can be graphically illustrated in the following figure
\begin{center}
\psfrag{u}{$u$}
\psfrag{u+}{$u+1$}
\psfrag{u-}{$u-1$}
\psfrag{a}{$\alpha_1$}
\psfrag{b}{$\beta_1$}
\psfrag{c}{$\alpha_2$}
\psfrag{d}{$\bar{\beta}_2$}
\psfrag{e}{$\bar{\beta}_3$}
\psfrag{f}{$\bar{\alpha}_3$}
\includegraphics[width=0.23\textwidth]{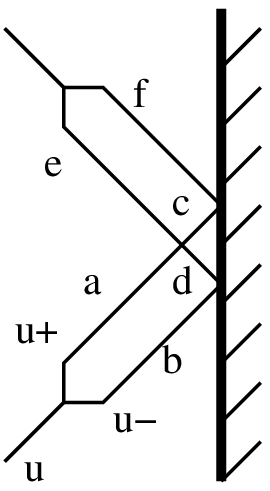}
\end{center}

If we do this calculation we get the following results:
\begin{align*}
	K_1^{(6)}(u) &= \begin{pmatrix}
					-1 & 0 & 0 & 0 & 0 & 0\\ 
					0 & -1 & 0 & 0 & 0 & 0\\ 
					0 & 0 & -1 & 0 & 0 & 0\\ 
					0 & 0 & 0 & 1 & 0 & 0\\ 
					0 & 0 & 0 & 0 & 1 & 0\\ 
					0 & 0 & 0 & 0 & 0 & 1 
				 \end{pmatrix}, &
	K_2^{(6)}(u) &= \begin{pmatrix}
					\frac{2+u}{2-u} & 0 & 0 & 0 & 0 & 0\\ 
					0 & 1 & 0 & 0 & 0 & 0\\ 
					0 & 0 & 1 & 0 & 0 & 0\\ 
					0 & 0 & 0 & 1 & 0 & 0\\ 
					0 & 0 & 0 & 0 & 1 & 0\\ 
					0 & 0 & 0 & 0 & 0 & 1 
				 \end{pmatrix}.
\end{align*}
We can see these two matrices are the same as the possible K-matrices of the \ref{Dn_d} case.

\section{Conclusion}

In this paper we developed a novel algebraic Bethe Ansatz method for open $\On{N}$ spin chains.
The two and three magnon eigenstates were calculated.
We have proposed the $n$-magnon eigenstate also.

The main result is the determination of Bethe Ansatz equations for $\On{2N}$ spin chains with $\On{2M+1}\times\On{2N-2M-1}$ reflection symmetry.
These open spin chains are interesting because the boundary breaks the rank of symmetry algebra so the number of Bethe roots is less than in the periodic case.
Based on these such models resemble $\mathrm{SU}(N)$ spin chains with non-soliton-preserving reflection.
Especially the $\On{6}$ models with $\On{5}$ or $\On{3}\times\On{3}$ symmetry are equivalent to the two non-soliton-preserving reflections of the $\mathrm{SU}(4)$ spin chain.

In the periodic case, the transfer matrix eigenvectors were defined by a complicated recursion rule.
We have seen that the multi-particle creation operator is not analogous to what was used in a periodic case.

The algebraic BA method developed in this paper can be generalized for the 
Sp(n) and Osp(n|2m) symmetric spin chains as in the periodic case.

\addtocontents{toc}{\protect\setcounter{tocdepth}{1}}

\section*{Acknowledgement}

I thank Zolt\'an Bajnok and L\'aszl\'o Palla for the useful discussions and for reading the manuscript.
The work was supported by a Lend\"ulet and by the NKFIH 116505 Grant.

\appendix

\section{Nesting for $\On{2N+1}$ model with $\On{2M}\times\On{2N-2M+1}$ boundary } 

The calculation which was introduced in subsection \ref{subsec:nesting} can be used in the $\On{2N+1}$ case also.
The R-matrix is the following:
\[
	R_{12}(u) = I_{12} - \frac{2}{u} P_{12} - \frac{2}{2N-1-u} K_{12} = 
	I_{12} + d(u) P_{12} + e_N(u) K_{12}.
\]
We can use the formulas of subsection \ref{subsec:nesting} with this new $e_N(u)$ and $\hat{u}=2N-1-u$ notation. The only difference is the last step of the nesting which is the $k=N-1$.

This $D_{N-1}$ is a transfer matrix of the O$(3)$ model which can be diagonalized with the same process which was used in the previous steps of the nesting.
At this step we have three dimensional quantum space.
We introduce a new basis: $\{\ket{2N-1},\ket{2N},\ket{2N+1}\}\longrightarrow\{\ket{N}_c,\ket{0},\ket{\bar{N}}_c\}$ where
\begin{align*}
	\frac{1}{\sqrt{2}}\left(\ket{2N-1}+i\ket{2N}\right) \longrightarrow & \ket{N}_c,\\
	\frac{1}{\sqrt{2}}\left(\ket{2N-1}-i\ket{2N}\right) \longrightarrow & \ket{\bar{N}}_c, \\
	\ket{2N+1} \longrightarrow & \ket{0}, \\
\end{align*}
In this basis the K-matrix and the monodromy matrix have the following form:
\begin{align*}
	K^{(N-1)} &= \begin{pmatrix}
       Y^{(N-1)} & 0 & 0 \\
	   0 & Y_0^{(N-1)} & 0 \\
	   0 & 0 & Y^{*(N-1)}
	  \end{pmatrix},&
	M^{(N-1)} &= \begin{pmatrix}
       A^{(N-1)} & B_0^{(N-1)} & B^{(N-1)} \\
	   C_0^{*(N-1)} & A_0^{(N-1)} & B_0^{*(N-1)} \\
	   C^{(N-1)} & C_0^{(N-1)} & A^{*(N-1)}
	  \end{pmatrix},
\end{align*} 
The calculation of section \ref{subsec:nesting} can be done in an analogous way, the difference is that $\Ym$, $\Am$ and $\Bv^t$ are replaced by $Y_0^{(N-1)}$, $A_0^{(N-1)}$ and $B_0^{(N-1)}$.
For example, the commutation relations of $A$s and $B_0$ are
\begin{align*}
  A^{(N-1)}(\theta) B_0^{(N-1)}(u) &= \frac{a(u-\theta)}{a(u+\theta-2N+2)} B_0^{(N-1)}(u) A^{(N-1)}(\theta) + \dots,\\
  \bar{A}_0^{(N-1)}(\theta) B_0^{(N-1)}(u) &= B_0^{(N-1)}(u)\bar{A}_0^{(N-1)}(\theta) R^{(N)}(\theta+u-2N)  R^{(N)}(\theta-u) + \dots ,\\
	\bar{A}^{(N-1)}(\theta) B_0^{(N-1)}(u) &= \frac{a(u-\hth)}{a(u+\hth-2N+2)} B_0^{(N-1)}(u) \bar{A}^{(N-1)}(\theta) + \dots,
\end{align*}
where $R^{(N)}(u)=1+d(u)+e_0(u)$.
The action of the $A$s on the pseudo-vacuum is:
\begin{align*}
	A_N^{(N-1)}(\theta)\ket{N}_c^{\otimes n_{N-1}} &= k_N^{R}(\theta)
	\prod_{i=1}^{n_{N-1}} a(\theta-\ui{N-1}{i}) a(\theta+\ui{N-1}{i})
	\ket{N}_c^{\otimes n_{N-1}}, \\
	A_0^{(N-1)}(\theta)\ket{N}_c^{\otimes n_{N-1}} &= \left[ Y^{(N-1)R}_{0}(\theta-N+1) -
	\frac{d(2\theta-2N+2)}{a(2\theta-2N+2)} Y^{(N-1)R}(\theta-N+1) \right]
	\ket{N}_c^{\otimes n_{N-1}}, \\
	A_{\bar{N}}^{(N-1)}(\theta)\ket{N}_c^{\otimes n_{N-1}} &=
	\bar{k}_{N}^{R}(\theta)  
	\prod_{i=1}^{n_{N-1}} a(\hth-\ui{N-1}{i}) a(\hth+\ui{N-1}{i})
	\ket{N}_c^{\otimes n_{N-1}}.
\end{align*}
The eigenvalue of $D^{(N-1)}(\theta)$ can be written in the following form:
\[
	\lambda_{N-1}(\theta) = F_{N-1}(\theta) + \lambda_N(\theta) + \bar{F}_{N-1}(\theta), 
\]
where 
\[
	\lambda_{N}(\theta) = k_N^L(\hth)k_N^R(\theta) 
	\prod_{i=1}^{n_{N}}
	\frac{\theta-\ui{N}{i}+2}{\theta-\ui{N}{i}}
	\frac{\theta+\ui{N}{i}-2N+2}{\theta+\ui{N}{i}-2N}
	\frac{\theta-\ui{N}{i}-1}{\theta-\ui{N}{i}+1}
	\frac{\theta+\ui{N}{i}-2N-1}{\theta+\ui{N}{i}-2N+1},
\]
where
\begin{align*}
	k_N^L(\theta) =	Y^{(N-1)L}_{0}(\theta-N+1) -
	\frac{d(2\theta-2N+2)}{a(2\theta-2N+2)} Y^{*(N-1)L}(\theta-N+1), \\
	k_N^R(\theta) = Y^{(N-1)R}_{0}(\theta-N+1) -
	\frac{d(2\theta-2N+2)}{a(2\theta-2N+2)} Y^{(N-1)R}(\theta-N+1).
\end{align*}

\subsection{$\On{2M}\times\On{2N-2M+1}$ symmetric boundaries }
In this subsection we use the following K-matrix:
\[
 K(u) = \mathrm{diag}( \underbrace{c(u),\dots,c(u)}_{2M}, \underbrace{1,\dots,1}_{2N-2M+1} ),
\]
where
\[
 c(u) = \frac{N+1/2-2M+u}{N+1/2-2M-u}.
\]
At the $k$th nesting step for $k<M$ the K-matrix is the following:
\[
 K^{(k)}(u) =\frac{u+k}{u} \frac{N+1/2-2M+k-u}{N+1/2-2M-k-u} \mathrm{diag}( \underbrace{c_k(u),\dots,c_k(u)}_{2M-2k}, \underbrace{1,\dots,1}_{2N-2M+1} ),
\]
where
\[
 c_k(u) = \frac{N+1/2-2M+k+u}{N+1/2-2M+k-u}.
\]
if $k\ge M$ then
\[
 K^{(k)}(u) =\frac{u+k}{u} \frac{N+1/2-k-u}{N+1/2-2M-k-u} \mathrm{diag}(\underbrace{1,\dots,1}_{2N-2k+1}) .
\]
The explicit form of the boundary coefficients for $k< M$:
\begin{align*}
  k_k^{L}(\theta) =& \bar{k}_k^{R}(\theta)= -\frac{(\hth-N+3/2)(\hth+N+1/2-2M)(\hth-2N+1)}{(\hth-k-1)(\hth-N+1/2)(\hth-N-2M+3/2)},\\
  \bar{k}_k^{L}(\theta) =& k_k^{R}(\theta) = - \frac{\theta(\theta+N+1/2-2M)}{(\theta-k)(\theta-N-1/2+2M)},
 \end{align*}
 and for $k\ge M$
\begin{align*}
  k_k^{L}(\theta) =& \bar{k}_k^{R}(\theta) = \frac{(\hth-N+3/2)(\hth-N-1/2)(\hth-2N+1)}{(\hth-k-1)(\hth-N+1/2)(\hth-N-2M+3/2)},\\
  \bar{k}_k^{L}(\theta) =& k_k^{R}(\theta) = \frac{\theta(\theta-N-1/2)}{(\theta-k)(\theta-N-1/2+2M)},
 \end{align*}
We still need the the boundary coefficients of $\lambda_N$.
For $M<N-1$
\begin{align*}
	k_N^L(\theta) = k_N^R(\theta) = \frac{\theta(\theta-N-1/2)}{(\theta-N)(\theta-N-1/2+2M)}
\end{align*}
and for $M=N$
\begin{align*}
	k_N^L(\theta) = k_N^R(\theta) = \frac{\theta(\theta-N-1/2)}{(\theta-N)(\theta+N-1/2)}
\end{align*}
It is convenient to redefine the rapidity variables:
\begin{align*}
 \vi{k}{i} &= \ui{k}{i} -k, \text{ where } k=1,\dots,N, &
\end{align*}
The BAEs are the following:
\begin{align*}
	r_1(\vi{1}{j})^2 \left( \frac{\vi{1}{j}-1}{\vi{1}{j}+1} \right)^{2L} 
	\prod_{k} \prod_{\substack{i=1\\(i,k)\neq(j,1)}}^{n_k}   
	s_{k1}(\vi{1}{j}-\vi{k}{i}) s_{k1}(\vi{1}{j}+\vi{k}{i}) = 1
\end{align*}
for $l=1$ and
\begin{align*}
	r_l(\vi{l}{j})^2 \prod_{k} \prod_{\substack{i=1\\(i,k)\neq(j,l)}}^{n_k}   
	s_{kl}(\vi{l}{j}-\vi{k}{i}) s_{kl}(\vi{l}{j}+\vi{k}{i}) = 1
\end{align*}
for $l=2,\dots,N$.
where the $\alpha_i$s in the scattering phase are the $B_N$ simple roots.
\[
	\alpha_i \cdot \alpha_j = 
	\begin{pmatrix}
		 2 & -1 &    &        &    &    & \\
		-1 &  2 & -1 &        &    &    & \\
		   & -1 &  2 &        &    &    & \\
		   &    &    & \ddots &    &    & \\
		   &    &    &        &  2 & -1 & \\
		   &    &    &        & -1 &  2 & -1  \\
		   &    &    &        &    & -1 &  1 \\
	\end{pmatrix}
\]
Only one reflection factor is non-trivial:
\[
 r_M(u) = \frac{N-M+\frac{1}{2}+u}{N-M+\frac{1}{2}-u}.
\]

\begin{center}
\psfrag{0}{$0$}
\psfrag{1}{$1$}
\psfrag{2}{$2$}
\psfrag{L}{$M$}
\psfrag{Nm1}{$N-1$}
\psfrag{N}{$N$}
\includegraphics[width=0.6\textwidth]{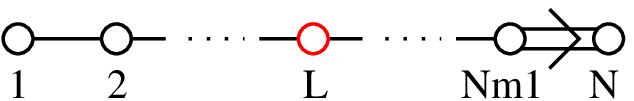}
\end{center}

\subsection{$\On{2}\times\On{2N-2M-1}$ symmetric boundaries }
In this subsection we use the following K-matrix:
\[
  K(u) = 
  \begin{pmatrix}
	\frac{(N-3/2)^2-c^2-u^2}{(N-3/2-u)^2-c^2} & i\frac{2cu}{(N-2-u)^2-c^2} & & & \\
	-i\frac{2cu}{(N-2-u)^2-c^2} &  \frac{(N-3/2)^2-c^2-u^2}{(N-3/2-u)^2-c^2} & & & \\
		&    & 1        &   \\
		&    & & \ddots &   \\
		&    & &        & 1 \\
  \end{pmatrix},
\]
where $K(u)$ is $2N+1\times2N+1$ dimensional matrix.
Now we select the matrices as in the previous case: $K^R=(K^L)^T=K$.
At the $k$th nesting step the K-matrix is the following:
\[
 K^{(k)}(u) =\frac{u+k}{u} \frac{N+1/2-k-c-u}{N-3/2-k-c-u} \mathrm{diag}(\underbrace{1,\dots,1}_{2N-2k+1}) .
\]
The explicit form of the boundary coefficients for $k=0$:
\begin{align*}
  k_0^{L}(\theta) =& \bar{k}_0^{R}(\theta) = -\frac{(\hth-N+3/2-c)(\hth+N-3/2-c)(\hth-2N+1)(\hth-n-1/2)}{(\hth-1)(\hth-N+1/2)(\hth-N+c-1/2)(\hth-N-c-1/2)},\\
  \bar{k}_0^{L}(\theta) =& k_0^{R}(\theta) = - \frac{\theta+N-3/2-c}{\theta-N+3/2+c},
 \end{align*}
and for $k>0$:
\begin{align*}
  k_k^{L}(\theta) =& \bar{k}_k^{R}(\theta) = \frac{(\hth-N+3/2-c)(\hth-N-1/2)(\hth-2N+1)}{(\hth-k-1)(\hth-N+1/2)(\hth-N-c-1/2)},\\
  \bar{k}_k^{L}(\theta) =& k_k^{R}(\theta) = \frac{\theta(\theta-N-1/2+c)}{(\theta-k)(\theta-N+3/2+c)},
 \end{align*}
 and
\begin{align*}
	k_N^L(\theta) = k_N^R(\theta) = \frac{\theta(\theta-N-1/2)}{(\theta-N)(\theta-N-1/2+2M)}
\end{align*}
The Bethe Ansatz equations are the same as which were at the previous case but the only not-trivial reflection is
\[
  r_1(u) = \frac{N-1/2-c-u}{N-1/2-c+u}.
\]

\section{$\On4$ model with $\mathrm{SU}_D(2)$ symmetry} \label{o4sol}

The O(4) R-matrix is the following:
\begin{align*}
  R_{ab}(u) &= I_{ab} -\frac{2}{u}P_{ab} -\frac{2}{2-u}K_{ab}, & 
  K(u)&=\mathrm{diag}(\frac{1-u}{u},\frac{1-u}{u},\frac{1-u}{u},\frac{1+u}{u}).
\end{align*}
We can diagonalize the transfer matrix if we use the local isomorphism $\On4\cong \mathrm{SU}(2)\times \mathrm{SU}(2)$.

The $\On4$ R-matrix is factorized in the following form:
\[
 R(\theta) = \frac{\theta}{\theta-2}  r(\theta) \otimes r(\theta),
\]
where $r(\theta)$ is the $\mathrm{SU}(2)$ R-matrix:
\[
 r(\theta) = \mathbb{I} -\frac{2}{\theta} P,
\]
where $P$ is the permutation operator.
The reflection matrix in the $\mathrm{SU}(2)\times \mathrm{SU}(2)$ basis:
\[
 K(\theta) = r(2\theta) P.
\]
We can see that this reflection has $\mathrm{SU}_D(2)$ symmetry which is the diagonal subgroup of $\mathrm{SU}(2)\times \mathrm{SU}(2)$. 

The $\On4$ double row monodromy matrix can be written in the following form:
\begin{multline*}
 D(\theta) =\\
  \tr{\as,\ad} \left[ r_{\as,\ad}(4-2\theta) P_{\as\ad} \prod_{k=1}^n \left[ r_{\as k}(\theta-\theta_k) r_{\ad \dot{k}}(\theta-\theta_k) \right] r_{\as,\ad}(2\theta) P_{\as\ad} \prod_{k=n}^1 \left[ r_{\ad \dot{k}}(\theta+\theta_k) r_{\as k}(\theta+\theta_k) \right] \right]
\end{multline*}
where
\begin{align*}
  D_0(\theta) = \prod_{k=1}^n \frac{\theta-\theta_k}{\theta-\theta_k-2} \frac{\theta+\theta_k}{\theta+\theta_k-2}.
\end{align*}
If we use the cyclicality of the trace we get the following:
\begin{multline*}
  D(\theta) = \\
  D_0(\theta) \tr{\as,\ad} \left[ r_{\as,\ad}(4-2\theta) \prod_{k=1}^n \left[ r_{\as k}(\theta-\theta_k) r_{\ad \dot{k}}(\theta-\theta_k) \right] r_{\as,\ad}(2\theta) \prod_{k=n}^1 \left[ r_{\as \dot{k}}(\theta+\theta_k) r_{\ad k}(\theta+\theta_k) \right] \right],
\end{multline*}
Using the Yang-Baxter equation we get the following:
\begin{multline*}
 D(\theta) = \\
 D_0(\theta) \tr{\as,\ad} \left[ r_{\as,\ad}(4-2\theta) \prod_{k=1}^n \left[ r_{\as k}(\theta-\theta_k) r_{\as \dot{k}}(\theta+\theta_k) \right] r_{\as,\ad}(2\theta) \prod_{k=n}^1 \left[ r_{\ad \dot{k}}(\theta-\theta_k) r_{\ad k}(\theta+\theta_k) \right] \right]
\end{multline*}
We can introduce new notations.
\begin{align*}
 T_{\as}(\theta) &= \prod_{k=1}^n \left[ r_{\as k}(\theta-\theta_k) r_{\as \dot{k}}(\theta+\theta_k) \right], & \widehat{T}_{\ad}(\theta) &= \prod_{k=n}^1 \left[ r_{\ad \dot{k}}(\theta-\theta_k) r_{\ad k}(\theta+\theta_k) \right].\\
 D(\theta) &= D_0(\theta) \tr{\as,\ad} \left[ r_{\as,\ad}(4-2\theta) T_{\as}(\theta) r_{\as,\ad}(2\theta) \widehat{T}_{\ad}(\theta) \right]
\end{align*}
Using the explicit expression of $r$ we get the following equation:
\[
 \mathcal{D}(\theta) = D_0(\theta) \tau(\theta) \widehat{\tau}(\theta),
\]
where
\begin{align*}
 \tau(\theta) &= \tr{\as} \left[ T_{\as}(\theta) \right],  & \widehat{\tau}(\theta) &= \tr{\ad} \left[ \widehat{T}_{\ad}(\theta) \right].
\end{align*}
If we use the crossing property of $r(u)$
\[
  r_{\as,\ad}(u) = \frac{u-2}{u}
  \left(\begin{pmatrix}
	0 & -i \\
	i & 0
  \end{pmatrix}\otimes
  \begin{pmatrix}
	1 & 0 \\
	0 & 1
\end{pmatrix}\right)
  r_{\as,\ad}^{t_\as}(2-u)
  \left(\begin{pmatrix}
	0 & -i \\
	i & 0
  \end{pmatrix}\otimes
  \begin{pmatrix}
	1 & 0 \\
	0 & 1
\end{pmatrix}\right)
\]
we can show the following identity:
\[
  \prod_{k=1}^n \frac{\theta-\theta_k}{\theta-\theta_k-2} \frac{\theta+\theta_k}{\theta+\theta_k-2} \widehat{\tau}(\theta) =  \tau(2-\theta).
\]
Thus
\[
 D(\theta) = \tau(\theta) \tau(2-\theta),
\]
Here $\tau$ is an one row transfer matrix so the $\tau$s commute at different rapidities that is we have to diagonalize only $\tau(\theta)$ and this is equivalent to the periodic XXX spin-chain problem.
The eigenvalue of $\tau(\theta)$ is:
\[
 \lambda(\theta) = \prod_{i=1}^n \frac{\theta-\theta_i-2}{\theta-\theta_i} \frac{\theta+\theta_i-2}{\theta+\theta_i} \prod_{i=1}^m \frac{\theta-u_i+2}{\theta-u_i} + \prod_{i=1}^m \frac{\theta-u_i-2}{\theta-u_i},
\]
where
\[
 \prod_{i=1}^n \frac{u_j-\theta_i-2}{u_j-\theta_i}\frac{u_j+\theta_i-2}{u_j+\theta_i} = - \prod_{i=1}^m \frac{u_j-u_i-2}{u_j-u_i+2}.
\]
The eigenvalue of $\mathcal{D}(\theta)$ is the following:
\[
 \Lambda(\theta) =  \lambda(\theta) \lambda(2-\theta).
\]

\section{$\beta\gamma$ type commutation relations} \label{appenbc}

To find the effect of the $A$, $\Am$ and $A^*$ operators on the pseudo-vacuum, we need the commutation relation of $\gammav^*(u)\hbetav^t(u)$, $\gamma(u)\hbeta(u)$ and $\gammav_1^t(u)\hbetav_2^*(u)$.
For these we will use the following equation:
\begin{align} \label{eq:TRT}
	T_{1}(u) R_{12}(2u) \hat{T}_2(u) = \hat{T}_2(u) R_{12}(2u) T_{1}(u).
\end{align}
We will start with the $\gammav(u)^*\hbetav^t(u)$ commutation relation.
We will need two equations of \eqref{eq:TRT}.
\begin{align} 
	a(2u) \gammav_1^* \hbetav^t_2 &= \hbetav^t_2 \Xm_{12}(2u) \gammav^*_1 +
	d(2u) (\halpha \alpha + \hbeta \gamma)) \kvt2
	- d(2u) \alpham_{1} \kvt2 \halpham_{2}
	-c_N(2u) \betav_1^*\hgammav_2^t, \label{eq:gambet1a}\\
	a(2u) \alpha \halpha &= a(2u) \halpha \alpha +
	c_N(2u) \hbeta \gamma + d(2u) \hbetav^t \gammav^* - d(2u)\betav^t \hgammav^*
	-c_N(2u) \beta \gamma. \label{eq:gambet1b}
\end{align}
We focus only on those terms that do not annihilate the pseudo-vacuum.
The commutation relation is:
\begin{align} \label{eq:gambet1c}
	\gammav^*\hbetav^t &= \frac{d(2u)}{a(2u)} \alpha \halpha -
						\frac{d(2u)}{a(2u)} \alpham \halpham + \dots
\end{align}
Let's continue with the $\gamma(u)\hbeta(u)$ and $\gammav_1^t(u)\hbetav_2^*(u)$ commutation relation.
We will use four equations from eq. \eqref{eq:TRT}.
\begin{align}
	a(2u) \gamma \hbeta &= -d(2u) \gammav^t\hbetav^* - 
	c(2u)(\alpha^* \halpha^* - \halpha \alpha) + 
	d(2u) \hbetav^t\gammav^* + a(2u) \hbeta \gamma, \label{eq:gambet2a}\\
	\gammav_1^t \Xm_{12}(2u) \hbetav_2^* &= -d(2u) (\gamma \hbeta \kvt1 + 
	\alpha^* \halpha^* \kvt1 - \halpham_{2} \kvt1 \alpham_{1}) + 
	c_N(2u) \hgammav^*_2\betav^t_1 + a(2u) \hbetav_2^*\gammav_1^t, \label{eq:gambet2b}\\
	d(2u) \halpham_{2} \kvt1 \alpham_{1} &= -a(2u) \hgammav_2^*\betav_1^t - 
	c_N(2u) \hbetav_2^*\gammav_1^t + \betav_1^t \Xm_{12}(2u) \hgammav_2^* + 
	d(2u) (\alpha \halpha + \beta \hgamma) \kvt1, \label{eq:gambet2c}\\
	\halpham_{2} \Xm_{12}(2u) \alpham_{1} &= -d(2u) (\hgammav_2^* \kvt2 \betav_1^t + 
	\hbetav_2^* \kvt2 \gammav_1^t) + d(2u) (\gammav_1^* \kvt1 \hbetav_2^t + 
	\betav_1^* \kvt1 \hgammav_2^t) + \alpham_{1} \Xm_{12}(2u) \halpham_{2}. \label{eq:gambet2d}
\end{align}
The products $\halpham_{2}\alpham_{1}$ and $\hgammav^*\betav^t$ can be expressed from the \eqref{eq:gambet2c}, \eqref{eq:gambet2d} and \eqref{eq:gambet1c} equations:
\begin{align}
	\halpham_{2} R_{12}^{(1)}(2u-2) \alpham_{1}&=
	\alpham_{1} R_{12}^{(1)}(2u-2) \halpham_{2}+ \dots \\
	\halpham_1 \alpha_1 &= \frac{1}{f(2u)}
				\mathrm{tr}_2 \left( \alpham_{2} R_{12}^{(1)}(2u-2)P_{12} \halpham_{2} \right)
	+ \dots \label{eq:gambet2e}\\
	\hgammav_1^* \betav_1^t &= -\frac{1}{f(2u)} \frac{d(2u)}{a(2u)}
	\mathrm{tr}_2 \left( \alpha_{2} R_{12}^{(1)}(2u-2)P_{12} \halpha_{2} \right) +
	\frac{d(2u)}{a(2u)}	\alpha \halpha I_1 + \dots \label{eq:gambet2f}
\end{align}
where
\[
	f(2u) = 1 + (2N-2) \frac{d(2u)}{a(2u)} + e_N(2u) = 1 + (2N-2)d(2u-2) + e_{N-1}(2u-2).
\]
From \eqref{eq:gambet2a}, \eqref{eq:gambet2b}, \eqref{eq:gambet2e}, \eqref{eq:gambet2f} we can get the expression of $\gamma(u)\hbeta(u)$ and $\gammav^t(u)\hbetav^*(u)$.
\begin{align*}
	\gamma \hbeta &= \frac{c_N(2u)}{a(2u)} \alpha \halpha + \frac{d(2\hat{u})}{a(2\hat{u})}
	\frac{d(2u)}{a(2u)} \mathrm{tr} \left( \alpham \halpham \right) +
	\frac{c_N(2\hat{u})}{a(2\hat{u})} \alpha^* \halpha^* + \dots \\
	\gammav_1^t\hbeta_2^* &= 
	\frac{d(2\hat{u})}{a(2\hat{u})} \left[ \alpha^*\halpha^* 
	- \mathrm{tr} \left( \alpha_{3} \underline{k}_{13} \underline{k}_{23}^{t_2t_3} \halpha_{3} \right) \right] + \dots
\end{align*}

\section{Commutation relations of $A$s with $B$} \label{appenAB}
To calculate the commutation relations we will use the following equation:
\begin{align} \label{eq:RMRM}
	R_{12}(u-v) M_1(u) R_{21}(u+v) M_2(v) = M_2(v) R_{12}(u+v) M_1(u) R_{21}(u-v).
\end{align}
One of the above equations is the following:
\begin{multline} \label{eq:comA1}
	\Bv^t(u) A(\theta) a(u-\theta) + B(u)\Cv^{*t}(\theta)a(u-\theta)c_N(u+\theta) = \\ 
	A(\theta)\Bv^t(u)a(u+\theta) + \Bv^t(\theta)\Am(u)d(u+\theta) + 
	B(\theta)\Cv^t(u)c_N(u+\theta) +  \\ 
	+ \Bv^t(\theta)A(u)d(u-\theta) + B(\theta)\Cv^{*t}(u)e_N(u+\theta)d(u-\theta).
\end{multline}
We can see that this is the commutation relation $A\Bv^t$:
\begin{multline} \label{eq:comAB1}
  A(\theta)\Bv^t(u) = \frac{a(u-\theta)}{a(u+\theta)} \Bv^t(u)A(\theta) - \frac{d(u-\theta)}{a(2u)} \Bv^t(u)A(\theta) - \frac{d(u+\theta)}{a(u+\theta)} \Bv^t(u)\bar{\Am}(\theta) + \dots 
\end{multline}
where $\dots$ denotes the $B\Cv^t$ type ''easy unwanted'' terms.

To calculate the commutation relation $\Am\Bv^t$ we will need two equations of \eqref{eq:RMRM}:
\begin{multline} \label{eq:comAab1}
	\Am_1(\theta)\Bv_2^t(u) + \Bv_1^*(\theta)\kvtt \Am_{2}e_N(\theta+u) +
	I_1 A(\theta)\Bv_2^t(u)d(\theta-u)d(\theta+u) + \\
	 + \kvt2 \Bv_1(\theta) \Xm_{12}(\theta+u) \Am_{2}(u)d(\theta-u)+  
	I_1 B(\theta) \Cv_2^t(u)d(\theta-u)d(\theta+u) = \\
	 \Bv_2^t(u)\Xm_{12}(\theta+u)\Am_{1}(\theta)\Xm_{12}(\theta-u) + 
	A(u) \kvt2 \Bv_1^t(\theta)\Xm_{12}(\theta-u)d(\theta+u) + \\ 
	 +B(u) \kvt2 \Cv_1^t(\theta)\Xm_{12}(\theta-u)d(\theta+u) +  
	B(u) \Cv_1^*(\theta) \kvtt e_N(\theta-u) + \\
	+ \Bv_1(u)A(\theta) \kvtt e_N(\theta-u)e_N(\theta+u) +\\
	+ (A(u)\Bv_1^*(\theta) +
	\Bv_1(u)A^*(\theta)e_N(\theta+u)) \kvtt e_N(\theta-u)
\end{multline}
and
\begin{multline} \label{eq:comAab2}
	\Bv_1^*(\theta)A(u)b_N(\theta+u) + \Bv_1(\theta)A(u)d(\theta-u)e_N(\theta+u) +
	B(\theta)\Cv_1^*(u)d(\theta-u) = \\
	\Bv_2^t(u)\Xm_{12}(\theta+u)\Am_{1}(\theta)\kv e_N(\theta-u) + 
	A(u)\Bv_1(\theta)e_N(\theta-u)d(\theta+u) + \\
	+ B(u)\Cv_1(\theta)e_N(\theta-u)d(\theta+u) + B(u)\Cv_1^*(\theta)c_N(\theta-u) + \\
	+ \Bv_1(u)A(\theta)c_N(\theta-u)e_N(\theta+u) +\\
	+ (A(u)\Bv_1^*(\theta) +
	\Bv_1(u)A^*(\theta)e_N(\theta+u))b_N(\theta-u).
\end{multline}
From \eqref{eq:comA1}, \eqref{eq:comAab1} and \eqref{eq:comAab2} we can express $A_{ab}B_c$.
\begin{multline}
	\Am_{1}(\theta)\Bv_2^t(u) = 
	\Bv_2^t(u) R_{12}^{(1)}(\theta+u-2) \Am_{1}(\theta) R_{12}^{(1)}(\theta-u) + \\
	+\Bv_1(u) A(\theta) \kvtt e(\theta+u) \left( e_N(\theta-u) -
	\frac{c_N(\theta-u)e_N(\theta-u)}{b_N(\theta-u)} \right) - \\
	-\kvt2 \Bv_1^t(u) A(\theta) R_{12}^{(1)}(\theta-u) d(\theta-u)
	\frac{d(\theta+u)}{a(\theta+u)} -
	I_1 \Bv_2^t(u) A(\theta) \delta_{ab} d(\theta-u)a(u-\theta)
	\frac{d(\theta+u)}{a(\theta+u)} - \\ - \Bv_1^*(\theta)\kvtt \Am_{2}e_N(\theta+u) +
    \Bv_1^*(\theta)A(u)\kvtt b_N(\theta+u) \frac{e_N(\theta-u)}{b_N(\theta-u)} - \\
	- \kvt2 \Bv_1(\theta) R_{12}^{(1)}(\theta+u-2) \Am_{2}(u) d(\theta-u) + 
	I_1 \Bv_2^t(\theta)\Am_2(u)d(\theta-u)\frac{d(\theta+u)}{a(\theta+u)} +\\
	+ \Bv_1(\theta)A(u)\kvtt d(\theta-u)e_N(\theta+u) \frac{e_N(\theta-u)}{b_N(\theta-u)}
	I_1 \Bv_2^t(\theta)A(u)d(\theta+u)\frac{d(\theta-u)}{a(\theta+u)} +\\
	+ \kvt2 \Bv_1^t(\theta) A(u) R_{12}^{(1)}(\theta-u)d(\theta+u)
	\frac{a(\theta-u)}{a(\theta+u)} + \dots
\end{multline}
where $\dots$ denotes the ''easy unwanted'' terms.
We can cancel the $\Bv(u)A(\theta)$ type ''wanted'' term if we use definition \eqref{eq:defAab}:
\begin{multline} \label{eq:comAB2}
	\bar{\Am}_{1}(\theta)\Bv_2^t(u) = 
	\Bv_2^t(u) R_{12}^{(1)}(\theta+u-2) \bar{\Am}_{1}(\theta) R_{12}^{(1)}(\theta-u) -\\
	- \Bv_1^*(\theta)\kvtt \bar{\Am}_{2}e_N(\theta+u) +
	\Bv_1^*(\theta)A(u)\kvtt \frac{1}{a(2u)} \frac{e_N(\theta-u)}{b_N(\theta-u)} - \\
	- \kvt2 \Bv_1(\theta) R_{12}^{(1)}(\theta+u-2) \bar{\Am}_{2}(u) d(\theta-u) + 
	I_1 \Bv_2^t(\theta)\bar{\Am}_2(u)
					\frac{d(2\theta)}{a(2\theta)} \frac{d(\theta+u)}{a(\theta+u)} +\\
	+ \Bv_1(\theta)A(u)\kvtt \frac{d(\theta+u)}{a(\theta+u)} \frac{e_N(\theta-u)}{a(2u)} +
	I_1 \Bv_2^t(\theta)A(u) \frac{1}{a(2u)}
					\frac{d(2\theta)}{a(2\theta)} \frac{d(\theta+u)}{a(\theta+u)} +\\
	+ \kvt2 \Bv_1^t(\theta) A(u) \frac{1}{a(2u)}  \frac{d(\theta+u)}{a(\theta+u)} + \dots
\end{multline}

Finally, let's look at the third commutation relation.
We will need one more equation from \eqref{eq:RMRM}.
\begin{multline} \label{eq:comAbegy}
	A^*(\theta)\Bv^t(u)b_N(\theta-u)b_N(\theta+u) +
	\mathrm{tr}\Am(\theta)\Bv^t(u)e_N(\theta-u)e_N(\theta+u) + \\ 
	+ \Bv^{*t}(\theta)\Am(u)e_N(\theta-u) + 
	B(\theta)\Cv^t(u)c_N(\theta-u)a(\theta+u) + \\
	+ \Bv^t(\theta)\Am(u)c_N(\theta-u)d(\theta+u) +
	A(\theta)\Bv^t(u)c_N(\theta-u)c_N(\theta+u) = \\
	\Bv^t(u)A^*(\theta) + A(u)\Bv^{*t}(\theta)e_N(\theta+u) +
	B(u)\Cv^t(\theta)d(\theta-u)a(\theta+u) + \\
	+ \Bv^t(u)\Am(\theta)d(\theta-u)d(\theta+u) +
	A(u)\Bv^t(\theta)d(\theta-u)c(\theta+u).
\end{multline}
If we use \eqref{eq:comA1}, \eqref{eq:comAab2}, \eqref{eq:comAB2}, \eqref{eq:comAbegy} and definitions \eqref{eq:defAab}, \eqref{eq:defAbegy} we get:
\[
	\bar{A}^*(\theta)\Bv^t(u) =
	\frac{a(u-\hth)}{a(u+\hth)} \Bv^t(u)\bar{A}^* + \dots
\]
where $\dots$ denote all of the ''unwanted'' terms.

\section{Two-particle states} \label{app:2part}
The two particle state is the following:
\[
	\ket{\Psi(u_1,u_2)}^{2-particle} = 
	\vec{\Phi}^{(2)}_{12}(u_1,u_2)\Fvec_{12}(u_1,u_2)\ket{1}_c^{\otimes L},
\]
where $\Fvec_{12}(u_1,u_2)$ an eigenvector of the $D^{(1)}_{12}(\theta)$ with $u_1$ and $u_2$ rapidities and
\[
  \vec{\Phi}^{(2)}_{12}(u_1,u_2) = \vec{B}_1(u_1)\vec{B}_2(u_2) -
  b_N(v_{12}) \frac{e_N(u_{12})}{b_N(u_{12})}\vec{k}_{12}B(u_1)A(u_2) +
  e_N(v_{12})\vec{k}_{12}B(u_1)\Am_{2}(u_2),
\]
or with $\bar{\Am}$ operator:
\[
  \vec{\Phi}^{(2)}_{12}(u_1,u_2) = \vec{B}_1(u_1)\vec{B}_2(u_2) -
  \frac{1}{a(2u_2)} \frac{e_N(u_{12})}{b_N(u_{12})}\vec{k}_{12}B(u_1)A(u_2) +
  e_N(v_{12})\vec{k}_{12}B(u_1)\bar{\Am}_{2}(u_2).
\]
This transfer matrix has the following symmetry:
\[
  D^{(1)}_{21}(\theta,u_2,u_1) = \frac{1}{a(u_{12})a(u_{21})} R_{12}^{(1)}(u_{21}) 
  D^{(1)}_{12}(\theta,u_1,u_2) R_{12}^{(1)}(u_{12}).
\]
It follows that
\[
  \Fvec_{21}(u_2,u_1) \sim \frac{1}{a(u_{21})} R_{12}^{(1)}(u_{21})\Fvec_{12}(u_1,u_2)
\]
If we want $\ket{\Psi(u_1,u_2)}^{2-particle}$ to be symmetric in the $u_1$ and $u_2$ variable, then the following condition is required for $\vec{\Phi}^{(2)}_{12}(u_1,u_2)$:
\begin{align} \label{eq:wavefunc2}
  \vec{\Phi}^{(2)}_{21}(u_2,u_1) = \frac{1}{a(u_{21})} \vec{\Phi}_{12}(u_1,u_2) R_{12}^{(1)}(u_{21}).
\end{align}
At this section we will prove that this condition is fulfilled.

We need two equations of \eqref{eq:RMRM}.
\begin{multline} \label{eq:2part1}
  B(u_2)A(u_1) a(u_{21})b_N(v_{12}) = \\
				\left( \vec{B}_1(u_1)\vec{B}_2(u_2) +
				\vec{k}_{12} B(u_1)\Am_{2}(u_2) e_N(v_{12}) \right)\kv e_N(u_{21})+\\
  +\left( A(u_1)B(u_2)a(v_{12})  + B(u_1)A^*(u_2) c_N(v_{12}) + 
  B(u_1)A(u_2) b_N(v_{12}) \frac{c_N(u_{21})}{b_N(u_{21})} \right) b_N(u_{21}),
\end{multline}
and
\begin{multline} \label{eq:2part2}
  \left( \vec{B}_2(u_2)\vec{B}_1(u_1) + \vec{k}_{12}B(u_2)\Am_{1}(u_1) e_N(v_{12}) \right) a(u_{21}) = \\
  \left( \vec{B}_1(u_1)\vec{B}_2(u_2) + \vec{k}_{12}B(u_1)\Am_{2}(u_2) e_N(v_{12}) \right) R_{12}(u_{21}) + \\ 
  + \biggl( A(u_1)B(u_2)a(v_{12})  + B(u_1)A^*(u_2) c_N(v_{12}) + 
  B(u_1)A(u_2) b_N(v_{12}) \biggr) \vec{k}_{12} e_N(u_{21}),
\end{multline}
Using equation \eqref{eq:2part1} and \eqref{eq:2part2} then we get the equation \eqref{eq:wavefunc2}.

\section{Three-particle states} \label{app:3part}
The tree-particle state is the following:
\[
	\ket{\Psi(u_1,u_2,u_3)}^{3-particle} = 
	\vec{\Phi}^{(3)}_{123}(u_1,u_2) \Fvec_{123}(u_1,u_2,u_3) 
	\ket{1}_c^{\otimes L},
\]
where
\begin{align*} 
  \vec{\Phi}^{(3)}_{123}(u_1,u_2,u_3) = \vec{B}_1(u_1)\vec{\Phi}^{(2)}_{23}(u_2,u_3) -&\\
  - \frac{a(u_{32})}{a(v_{32})} \frac{1}{a(2u_2)} \frac{e_N(u_{12})}{b_N(u_{12})}
  & \vec{k}_{12}B(u_1)\vec{\Phi}^{(1)}_3(u_3)A(u_2) -\\
  - \frac{1}{a(v_{32})} \frac{1}{a(2u_3)} \frac{e_N(u_{13})}{b_N(u_{13})}
  & \vec{k}_{13}B(u_1)\vec{\Phi}^{(1)}_2(u_2)A(u_3) R_{23}^{(1)}(u_{23}) +\\
  + e_N(v_{12}) & \vec{k}_{12}B(u_1)\vec{\Phi}^{(1)}_3(u_3)
  R_{32}^{(1)}(v_{32}-2) \bar{\Am}_{2}(u_2) R_{23}^{(1)}(u_{23}) + \\
  + a(u_{32}) e_N(v_{13}) & \vec{k}_{13}B(u_1)\vec{\Phi}^{(1)}_2(u_2)
  R_{23}^{(1)}(v_{23}-2) \bar{\Am}_{3}(u_3) .
\end{align*}
This eigenvector has to have the following symmetries:
\begin{align}
	\vec{\Phi}^{(3)}_{213}(u_2,u_1,u_3)\ket{1}_c^{\otimes L} &= 
	\frac{1}{a(u_{21})} \vec{\Phi}^{(3)}_{123}(u_1,u_2,u_3) 
	R_{12}^{(1)}(u_{21})\ket{1}_c^{\otimes L}, \label{eq:wavefunc31}\\
	\vec{\Phi}^{(3)}_{132}(u_1,u_3,u_2)\ket{1}_c^{\otimes L} &= 
	\frac{1}{a(u_{32})} \vec{\Phi}^{(3)}_{123}(u_1,u_2,u_3) 
	R_{23}^{(1)}(u_{32})\ket{1}_c^{\otimes L}. \label{eq:wavefunc32}
\end{align}
At this section we will prove that this condition is fulfilled.

Equation \eqref{eq:wavefunc32} follows from \eqref{eq:wavefunc2}.
To prove equation \eqref{eq:wavefunc31} we have to use \eqref{eq:comAB1} and \eqref{eq:comAB2} commutation relations to change the order of the rapidities to $u_1,u_2,u_3$ for each term. The result is the following:
\begin{align} \label{eq:wave3m}
\begin{split} 
  \vec{\Phi}^{(3)}_{123}(u_1,u_2,u_3)\ket{1}_c^{\otimes L} =
  \Biggl\{ \vec{B}_1(u_1)\vec{\Phi}^{(2)}_{23}(u_2,u_3) -&\\
  - \frac{1}{a(2u_2)} \frac{e_N(u_{12})}{b_N(u_{12})}
  & \vec{k}_{12}B(u_1)A(u_2)\vec{B}_3(u_3) +\\
  + e_N(v_{12}) & \vec{k}_{12}B(u_1)\bar{\Am}_{2}(u_2)\vec{B}_3(u_3) - \\
  - \frac{e_N(v_{12})}{a(2u_3)} \frac{e_N(u_{23})}{b_N(u_{23})}
  & \vec{k}_{23} B(u_1) \vec{B}_1^*(u_2) A(u_3) +\\
  + e_N(v_{12}) e_N(v_{23}) & \vec{k}_{23} B(u_1)\vec{B}_1^*(u_2) \bar{\Am}_3(u_3)  - \\
  - \frac{b_N(v_{12})}{a(2u_3)} \frac{e_N(u_{13})}{b_N(u_{13})} \biggl[
	\frac{e_N(u_{23})}{b_N(u_{23})} & \vec{k}_{23} B(u_1) \vec{B}_1(u_2) A(u_3) +\\
								+	& \vec{k}_{13} B(u_1) \vec{B}_2(u_2) A(u_3) +\\
	+\frac{e_N(u_{12})}{b_N(u_{12})}& \vec{k}_{12} B(u_1) \vec{B}_3(u_2) A(u_3) \biggr] + \\
  + a(u_{32}) e_N(v_{13}) \biggl[
					e_N(v_{23}) & \vec{k}_{23} B(u_1) \vec{B}_1(u_2) \bar{\Am}_3(u_3) +\\
						+		& \vec{k}_{13} B(u_1) \vec{B}_2(u_2) \bar{\Am}_3(u_3) +\\
    +\frac{d(v_{23})}{a(v_{23})}& \vec{k}_{12} B(u_1) \vec{B}_3(u_2) \bar{\Am}_3(u_3) \biggr] \Biggr\}
	\ket{1}_c^{\otimes L}
\end{split}
\end{align}

We need two equations of \eqref{eq:RMRM}.
\begin{multline} \label{eq:3part1}
	a(u_{21})\left(\vec{B}(u_2)B(u_1) + e_N(v_{12})B(u_2)\vec{B}^*(u_1)\right) = \\
	d(u_{21})\left(\vec{B}(u_1)B(u_2) + e_N(v_{12})B(u_1)\vec{B}^*(u_2)\right) + 
	b_N(v_{12}) B(u_1)\vec{B}(u_2),
\end{multline}
and
\begin{multline} \label{eq:3part2}
	a(u_{21}) b_N(v_{12}) B(u_2)\vec{B}(u_1) = \\
	d(u_{21}) b_N(v_{12}) B(u_1)\vec{B}(u_2) +
	\left(\vec{B}(u_1)B(u_2) + e_N(v_{12})B(u_1)\vec{B}^*(u_2)\right).
\end{multline}
If we use  equations \eqref{eq:3part1}, \eqref{eq:3part2}, \eqref{eq:wavefunc2} and \eqref{eq:wave3m} then we can derive equation \eqref{eq:wavefunc31}.

We can generalize the formula from 3-particle states to $n$-particle states:
\begin{align} 
  \begin{split} \label{nparticle}
  &\vec{\Phi}^{(n)}_{1\dots n}(u_1,\dots,u_n) =
  \vec{B}_1(u_1)\vec{\Phi}^{(n-1)}_{2\dots n}(u_2,\dots,u_n) -\\
   & - \sum_{j=2}^n \frac{1}{a(2u_j)} \frac{e_N(u_{1j})}{b_N(u_{1j})}
   \left[ \prod_{k=2,k\neq j}^n \frac{a(u_{kj})}{a(v_{kj})} \right] \vec{k}_{1j}B(u_1) \times \\
   & \times 
  \vec{\Phi}^{(n-2)}_{2,\dots,\hat{\jmath},\dots,n}(u_2,\dots,\widehat{u_{j}},\dots,u_n)A(u_j)
  \prod_{k=2}^{\substack{j-1\\ \longrightarrow}} \frac{R_{kj}^{(1)}(u_{kj})}{a(u_{kj})} +\\
   & + \sum_{j=2}^n e_N(v_{1j}) 
  \left[\prod_{k=2,k\neq j}^n a(u_{kj}) \right] \vec{k}_{1j}B(u_1) \times \\
   & \times \vec{\Phi}^{(n-2)}_{2,\dots,\hat{\jmath},\dots,n}(u_2,\dots,\widehat{u_{j}},\dots,u_n)
  \prod_{k=2,k\neq j}^{\substack{n\\ \longrightarrow}} R_{kj}^{(1)}(v_{kj}-2) \bar{\Am}_{j}(u_j) 
  \prod_{k=j+1}^{\substack{n\\ \longleftarrow}} \frac{R_{jk}^{(1)}(u_{jk})}{a(u_{jk})},
\end{split}
\end{align}
where hats denote the missing terms.
It is easy to show that the above state fulfills the following conditions for $j>1$:
\begin{multline*}
	\vec{\Phi}^{(n)}_{1,\dots,j+1,j,\dots}(u_1\dots,u_{j+1},u_j,\dots) 
	\ket{1}_c^{ \otimes L} 
	=\\ 
	\frac{1}{a(u_{j+1}-u_j)} 
	\vec{\Phi}^{(n)}_{1,\dots,j,j+1,\dots}(u_1,\dots,u_j,u_{j+1},\dots) 
	R_{j,j+1}^{(1)}(u_{j+1}-u_j)\ket{1}_c^{\otimes L}.
\end{multline*}
However the remaining condition ($j=1$) is not proved so the formula 
\eqref{nparticle} for the $n$-particle eigenstate is just a conjecture for 
$n>3$.
That is why $n=4$ magnon states were checked numerically in the 
$\On{3}$ model. 

The table \ref{tab:O3} contains some 4 magnon solutions of the Bethe 
Ansatz equation at the $\On{3}$ symmetric boundary case when $L=5$.
The tables \ref{tab:O2L4} and \ref{tab:O2L5} show 4 magnon states for $L=4$ and 
$L=5$ chains at the $\On{2}$ symmetric boundary case where the reflection 
matrix is \[ K(u)= 
	 \begin{pmatrix}
		  \frac{1/2-u}{1/2+u} & 0 & 0 \\
		  0 & \frac{1/2-u}{1/2+u} & 0 \\
		  0 &          0          & 1  
	 \end{pmatrix}.
\]
We can check numerically that using the formula \eqref{nparticle} with these 
rapidities we get eigenvectors of the transfer matrix. 

\begin{table}
\centering

\begin{tabular}{|c|c|}
	$v_{1,2}/i$      & $v_{3,4}/i$  \\
	\hline
	$0.15505\pm0.500072i$ &  $1.40879\pm0.596627i$ \\
	\hline
	$0.16835\pm0.500319i$ & $0.790932\pm0.476179i$ \\
	\hline
	$0.336147\pm0.499739i$ & $1.38729\pm0.597299i$ 
\end{tabular}
\caption{$L=5$, $\On{3}$ symmetric boundary}
\label{tab:O3}

\vspace{1cm}
\begin{tabular}{|c|c|c|}
	$v_{1,2}/i$      & $v_{3}/i$ & $v_{4}/i$ \\
	\hline
	$0.651897\pm0.529008i$ &  $0.503048i$ & $1.50742i$ 
\end{tabular}
\caption{$L=4$, $\On{2}$ symmetric boundary}
\label{tab:O2L4}

\vspace{1cm}
\begin{tabular}{|c|c|c|c|}
	$v_{1}/i$ & $v_{2}/i$ & $v_{3}/i$ & $v_{4}/i$ \\
	\hline
	$0.581077$ &  $0.500009i$ & $1.50001i$  & $2.55144i$ \\
	\hline
	$1.45725$  &  $0.500003i$ & $1.500003i$ & $2.53386i$ \\
	\hline
	$0.499999839i$  &  $1.499999841i$ & $2.50153i$ & $3.78771i$
\end{tabular}
\caption{$L=5$, $\On{2}$ symmetric boundary}
\label{tab:O2L5}
\end{table}

\bibliography{on.bib}

\begin{thebibliography}{10}

\bibitem{Arnaudon:2004sd}
D.~Arnaudon, J.~Avan, N.~Crampe, A.~Doikou, L.~Frappat, and E.~Ragoucy,
  ``{General boundary conditions for the sl(N) and sl(M|N) open spin chains},''
  {\em J. Stat. Mech.}, vol.~0408, p.~P08005, 2004.

\bibitem{Arnaudon:2003gj}
D.~Arnaudon, J.~Avan, N.~Crampe, A.~Doikou, L.~Frappat, and E.~Ragoucy,
  ``{Classification of reflection matrices related to (super) Yangians and
  application to open spin chain models},'' {\em Nucl. Phys.}, vol.~B668,
  pp.~469--505, 2003.

\bibitem{Doikou:2000yw}
A.~Doikou, ``{Quantum spin chain with 'soliton nonpreserving' boundary
  conditions},'' {\em J. Phys.}, vol.~A33, pp.~8797--8808, 2000.

\bibitem{Moriconi:2001xz}
M.~Moriconi, ``{Integrable boundary conditions and reflection matrices for the
  O(N) nonlinear sigma model},'' {\em Nucl. Phys.}, vol.~B619, pp.~396--414,
  2001.

\bibitem{MacKay:2001bh}
N.~J. MacKay and B.~J. Short, ``{Boundary scattering, symmetric spaces and the
  principal chiral model on the half line},'' {\em Commun. Math. Phys.},
  vol.~233, pp.~313--354, 2003.

\bibitem{Arnaudon:2005rq}
D.~Arnaudon, N.~Crampe, A.~Doikou, L.~Frappat, and E.~Ragoucy, ``{Analytical
  Bethe ansatz for open spin chains with soliton non preserving boundary
  conditions},'' {\em Int. J. Mod. Phys.}, vol.~A21, pp.~1537--1554, 2006.

\bibitem{Gerrard:2017igy}
A.~Gerrard, N.~MacKay, and V.~Regelskis, ``{Nested algebraic Bethe ansatz for
  open spin chains with even twisted Yangian symmetry},'' 2017.

\bibitem{Short:2002yr}
B.~J. Short, ``{Boundary scattering in the SU(N) principal chiral model on the
  half line with conjugating boundary conditions},'' {\em Phys. Lett.},
  vol.~B547, pp.~257--268, 2002.

\bibitem{deVega:1986xj}
H.~J. de~Vega and M.~Karowski, ``{Exact Bethe Ansatz Solution of 0(2n)
  Symmetric Theories},'' {\em Nucl. Phys.}, vol.~B280, pp.~225--254, 1987.

\bibitem{Gombor:2015kdu}
T.~Gombor and L.~Palla, ``{Algebraic Bethe Ansatz for O(2N) sigma models with
  integrable diagonal boundaries},'' {\em JHEP}, vol.~02, p.~158, 2016.

\bibitem{Martins:1997wb}
M.~J. Martins and P.~B. Ramos, ``{The Algebraic Bethe ansatz for rational braid
  - monoid lattice models},'' {\em Nucl. Phys.}, vol.~B500, pp.~579--620, 1997.

\bibitem{2005PhLA}
W.~{Galleas} and M.~J. {Martins}, ``{Solution of the SU(N) vertex model with
  non-diagonal open boundaries},'' {\em Physics Letters A}, vol.~335,
  pp.~167--174, 2005.

\end{thebibliography}
\end{document}